%% file: mnras_template.tex
\documentclass[fleqn,usenatbib]{mnras}

\usepackage{newtxtext,newtxmath}

\usepackage[T1]{fontenc}

\DeclareRobustCommand{\VAN}[3]{#2}
\let\VANthebibliography\thebibliography
\def\thebibliography{\DeclareRobustCommand{\VAN}[3]{##3}\VANthebibliography}

\usepackage{graphicx}	
\usepackage{amsmath}	

\title[Observing 23 M-dwarfs with \textit{CHEOPS}]{The EBLM Project XI. Mass, radius and effective temperature measurements for 23 M-dwarf companions to solar-type stars observed with \textit{CHEOPS}\thanks{This study uses data from the Guaranteed Time Observation (GTO) \textit{CHEOPS} programme CH$\_$PR100037}\thanks{The raw and detrended photometric time-series data are available in electronic form at the CDS via anonymous ftp to cdsarc.u-strasbg.fr (130.79.128.5) or via http://cdsweb.u-strasbg.fr/cgi-bin/qcat?J/MNRAS/}\thanks{Based in part on observations collected at the Observatoire de Haute-Provence (CNRS), France.}\thanks{Based on observations made with the Southern African Large Telescope (SALT)}}

\author[M. I. Swayne et al.]{
M.I. Swayne$^{1}$\thanks{E-mail: m.i.swayne@keele.ac.uk}$^{\href{https://orcid.org/0000-0002-2609-3159}{\includegraphics[scale=0.5]{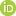}}}$, 
P.F.L. Maxted$^{1\,\href{https://orcid.org/0000-0003-3794-1317}{\includegraphics[scale=0.5]{figures/orcid.jpg}}}$, 
A.H.M.J. Triaud$^{2\,\href{https://orcid.org/0000-0002-5510-8751}{\includegraphics[scale=0.5]{figures/orcid.jpg}}}$,
S. G. Sousa$^{3\,\href{https://orcid.org/0000-0001-9047-2965}{\includegraphics[scale=0.5]{figures/orcid.jpg}}}$,
A. Deline$^{4}$, 
D. Ehrenreich$^{4,5\,\href{https://orcid.org/0000-0001-9704-5405}{\includegraphics[scale=0.5]{figures/orcid.jpg}}}$,\newauthor
S. Hoyer$^{6\,\href{https://orcid.org/0000-0003-3477-2466}{\includegraphics[scale=0.5]{figures/orcid.jpg}}}$, 
G. Olofsson$^{7\,\href{https://orcid.org/0000-0003-3747-7120}{\includegraphics[scale=0.5]{figures/orcid.jpg}}}$, 
I. Boisse$^{6}$,
A. Duck$^{8\,\href{https://orcid.org/0000-0002-4531-6899}{\includegraphics[scale=0.5]{figures/orcid.jpg}}}$,
S. Gill$^{9\,\href{https://orcid.org/0000-0002-4259-0155}{\includegraphics[scale=0.5]{figures/orcid.jpg}}}$, 
D. Martin$^{8\,\href{https://orcid.org/0000-0002-7595-6360}{\includegraphics[scale=0.5]{figures/orcid.jpg}}}$, 
J. McCormac$^{9}$, 
C.M. Persson$^{10}$,\newauthor
A. Santerne$^{6}$, 
D. Sebastian$^{2\,\href{https://orcid.org/0000-0002-2214-9258}{\includegraphics[scale=0.5]{figures/orcid.jpg}}}$, 
M.R. Standing$^{11\,\href{https://orcid.org/0000-0002-7608-8905}{\includegraphics[scale=0.5]{figures/orcid.jpg}}}$, 
L. Acuña$^{6\,\href{https://orcid.org/0000-0002-9147-7925}{\includegraphics[scale=0.5]{figures/orcid.jpg}}}$, 
Y. Alibert$^{12,13\,\href{https://orcid.org/0000-0002-4644-8818}{\includegraphics[scale=0.5]{figures/orcid.jpg}}}$, 
R. Alonso$^{14,15\,\href{https://orcid.org/0000-0001-8462-8126}{\includegraphics[scale=0.5]{figures/orcid.jpg}}}$, \newauthor
G. Anglada$^{16,17\,\href{https://orcid.org/0000-0002-3645-5977}{\includegraphics[scale=0.5]{figures/orcid.jpg}}}$, 
T. Bárczy$^{18\,\href{https://orcid.org/0000-0002-7822-4413}{\includegraphics[scale=0.5]{figures/orcid.jpg}}}$, 
D. Barrado Navascues$^{19\,\href{https://orcid.org/0000-0002-5971-9242}{\includegraphics[scale=0.5]{figures/orcid.jpg}}}$, 
S.C.C. Barros$^{3,20\,\href{https://orcid.org/0000-0003-2434-3625}{\includegraphics[scale=0.5]{figures/orcid.jpg}}}$, 
W. Baumjohann$^{21\,\href{https://orcid.org/0000-0001-6271-0110}{\includegraphics[scale=0.5]{figures/orcid.jpg}}}$, \newauthor
T.A. Baycroft$^{2\,\href{https://orcid.org/0000-0002-3300-3449}{\includegraphics[scale=0.5]{figures/orcid.jpg}}}$, 
M. Beck$^{4\,\href{https://orcid.org/0000-0003-3926-0275}{\includegraphics[scale=0.5]{figures/orcid.jpg}}}$, 
T. Beck$^{13}$, 
W. Benz$^{13,12\,\href{https://orcid.org/0000-0001-7896-6479}{\includegraphics[scale=0.5]{figures/orcid.jpg}}}$, 
N. Billot$^{4\,\href{https://orcid.org/0000-0003-3429-3836}{\includegraphics[scale=0.5]{figures/orcid.jpg}}}$, 
X. Bonfils$^{22\,\href{https://orcid.org/0000-0001-9003-8894}{\includegraphics[scale=0.5]{figures/orcid.jpg}}}$, 
L. Borsato$^{23\,\href{https://orcid.org/0000-0003-0066-9268}{\includegraphics[scale=0.5]{figures/orcid.jpg}}}$, \newauthor
V. Bourrier$^{4\,\href{https://orcid.org/0000-0002-9148-034X}{\includegraphics[scale=0.5]{figures/orcid.jpg}}}$, 
A. Brandeker$^{7\,\href{https://orcid.org/0000-0002-7201-7536}{\includegraphics[scale=0.5]{figures/orcid.jpg}}}$, 
C. Broeg$^{13,12\,\href{https://orcid.org/0000-0001-5132-2614}{\includegraphics[scale=0.5]{figures/orcid.jpg}}}$, 
A. Carmona$^{22\,\href{https://orcid.org/0000-0003-2471-1299}{\includegraphics[scale=0.5]{figures/orcid.jpg}}}$, 
S. Charnoz$^{24\,\href{https://orcid.org/0000-0002-7442-491X}{\includegraphics[scale=0.5]{figures/orcid.jpg}}}$, 
A. Collier Cameron$^{25\,\href{https://orcid.org/0000-0002-8863-7828}{\includegraphics[scale=0.5]{figures/orcid.jpg}}}$, \newauthor
P. Cortés-Zuleta$^{6\,\href{https://orcid.org/0000-0002-6174-4666}{\includegraphics[scale=0.5]{figures/orcid.jpg}}}$, 
Sz. Csizmadia$^{26\,\href{https://orcid.org/0000-0001-6803-9698}{\includegraphics[scale=0.5]{figures/orcid.jpg}}}$, 
P.E. Cubillos$^{27,21}$, 
M.B. Davies$^{28\,\href{https://orcid.org/0000-0001-6080-1190}{\includegraphics[scale=0.5]{figures/orcid.jpg}}}$, 
M. Deleuil$^{6\,\href{https://orcid.org/0000-0001-6036-0225}{\includegraphics[scale=0.5]{figures/orcid.jpg}}}$, 
X. Delfosse$^{22}$, \newauthor
L. Delrez$^{29,30\,\href{https://orcid.org/0000-0001-6108-4808}{\includegraphics[scale=0.5]{figures/orcid.jpg}}}$, 
O.D.S. Demangeon$^{3,20\,\href{https://orcid.org/0000-0001-7918-0355}{\includegraphics[scale=0.5]{figures/orcid.jpg}}}$, 
B.-O. Demory$^{12,13\,\href{https://orcid.org/0000-0002-9355-5165}{\includegraphics[scale=0.5]{figures/orcid.jpg}}}$, 
G. Dransfield$^{2\,\href{https://orcid.org/0000-0002-3937-630X}{\includegraphics[scale=0.5]{figures/orcid.jpg}}}$, 
A. Erikson$^{26}$, 
A. Fortier$^{13,12\,\href{https://orcid.org/0000-0001-8450-3374}{\includegraphics[scale=0.5]{figures/orcid.jpg}}}$, \newauthor
T. Forveille$^{22\,\href{https://orcid.org/0000-0003-0536-4607}{\includegraphics[scale=0.5]{figures/orcid.jpg}}}$, 
L. Fossati$^{21\,\href{https://orcid.org/0000-0003-4426-9530}{\includegraphics[scale=0.5]{figures/orcid.jpg}}}$, 
M. Fridlund$^{31,10\,\href{https://orcid.org/0000-0002-0855-8426}{\includegraphics[scale=0.5]{figures/orcid.jpg}}}$, 
D. Gandolfi$^{32\,\href{https://orcid.org/0000-0001-8627-9628}{\includegraphics[scale=0.5]{figures/orcid.jpg}}}$, 
M. Gillon$^{29\,\href{https://orcid.org/0000-0003-1462-7739}{\includegraphics[scale=0.5]{figures/orcid.jpg}}}$,
M. Güdel$^{33}$,  
M.N. Günther$^{34}$, \newauthor
N. Hara$^{4\,\href{https://orcid.org/0000-0001-9232-3314}{\includegraphics[scale=0.5]{figures/orcid.jpg}}}$, 
G. Hébrard$^{35,36}$, 
N. Heidari$^{35,37,6\,\href{https://orcid.org/0000-0002-2370-0187}{\includegraphics[scale=0.5]{figures/orcid.jpg}}}$, 
C. Hellier$^{1\,\href{https://orcid.org/0000-0002-3439-1439}{\includegraphics[scale=0.5]{figures/orcid.jpg}}}$, 
Ch. Helling$^{21,38}$, 
K.G. Isaak$^{34\,\href{https://orcid.org/0000-0001-8585-1717}{\includegraphics[scale=0.5]{figures/orcid.jpg}}}$, 
F. Kerschbaum$^{33}$, \newauthor
F. Kiefer$^{39,40}$, 
L.L. Kiss$^{41,42}$, 
V. Kunovac$^{9,43\,\href{https://orcid.org/0000-0001-9419-3736}{\includegraphics[scale=0.5]{figures/orcid.jpg}}}$, 
S. Lalitha$^{2\,\href{https://orcid.org/0000-0001-8102-3033}{\includegraphics[scale=0.5]{figures/orcid.jpg}}}$,
K.W.F. Lam$^{26\,\href{https://orcid.org/0000-0002-9910-6088}{\includegraphics[scale=0.5]{figures/orcid.jpg}}}$, 
J. Laskar$^{44\,\href{https://orcid.org/0000-0003-2634-789X}{\includegraphics[scale=0.5]{figures/orcid.jpg}}}$, \newauthor
A. Lecavelier des Etangs$^{35\,\href{https://orcid.org/0000-0002-5637-5253}{\includegraphics[scale=0.5]{figures/orcid.jpg}}}$, 
M. Lendl$^{4\,\href{https://orcid.org/0000-0001-9699-1459}{\includegraphics[scale=0.5]{figures/orcid.jpg}}}$, 
D. Magrin$^{23\,\href{https://orcid.org/0000-0003-0312-313X}{\includegraphics[scale=0.5]{figures/orcid.jpg}}}$, 
L. Marafatto$^{23}$, 
E. Martioli$^{45,35\,\href{https://orcid.org/0000-0002-5084-168X}{\includegraphics[scale=0.5]{figures/orcid.jpg}}}$, 
N.J. Miller$^{46\,\href{https://orcid.org/0000-0001-9550-1198}{\includegraphics[scale=0.5]{figures/orcid.jpg}}}$, \newauthor
C. Mordasini$^{13,12}$, 
C. Moutou$^{47\,\href{https://orcid.org/0000-0002-2842-3924}{\includegraphics[scale=0.5]{figures/orcid.jpg}}}$, 
V. Nascimbeni$^{23\,\href{https://orcid.org/0000-0001-9770-1214}{\includegraphics[scale=0.5]{figures/orcid.jpg}}}$, 
R. Ottensamer$^{33}$, 
I. Pagano$^{48\,\href{https://orcid.org/0000-0001-9573-4928}{\includegraphics[scale=0.5]{figures/orcid.jpg}}}$,
E. Pallé$^{14,15\,\href{https://orcid.org/0000-0003-0987-1593}{\includegraphics[scale=0.5]{figures/orcid.jpg}}}$, \newauthor
G. Peter$^{49\,\href{https://orcid.org/0000-0001-6101-2513}{\includegraphics[scale=0.5]{figures/orcid.jpg}}}$, 
D. Piazza$^{13}$,  
G. Piotto$^{23,50\,\href{https://orcid.org/0000-0002-9937-6387}{\includegraphics[scale=0.5]{figures/orcid.jpg}}}$, 
D. Pollacco$^{9}$, 
D. Queloz$^{51,52\,\href{https://orcid.org/0000-0002-3012-0316}{\includegraphics[scale=0.5]{figures/orcid.jpg}}}$, 
R. Ragazzoni$^{23,50\,\href{https://orcid.org/0000-0002-7697-5555}{\includegraphics[scale=0.5]{figures/orcid.jpg}}}$, 
N. Rando$^{34}$, \newauthor
H. Rauer$^{26,53,54\,\href{https://orcid.org/0000-0002-6510-1828}{\includegraphics[scale=0.5]{figures/orcid.jpg}}}$, 
I. Ribas$^{16,17\,\href{https://orcid.org/0000-0002-6689-0312}{\includegraphics[scale=0.5]{figures/orcid.jpg}}}$,  
N.C. Santos$^{3,20\,\href{https://orcid.org/0000-0003-4422-2919}{\includegraphics[scale=0.5]{figures/orcid.jpg}}}$, 
G. Scandariato$^{48\,\href{https://orcid.org/0000-0003-2029-0626}{\includegraphics[scale=0.5]{figures/orcid.jpg}}}$, 
D. Ségransan$^{4\,\href{https://orcid.org/0000-0003-2355-8034}{\includegraphics[scale=0.5]{figures/orcid.jpg}}}$, 
A.E. Simon$^{13\,\href{https://orcid.org/0000-0001-9773-2600}{\includegraphics[scale=0.5]{figures/orcid.jpg}}}$, \newauthor
A.M.S. Smith$^{26\,\href{https://orcid.org/0000-0002-2386-4341}{\includegraphics[scale=0.5]{figures/orcid.jpg}}}$, 
R. Southworth$^{34}$, 
M. Stalport$^{29,30}$, 
Gy.M. Szabó$^{55,56\,\href{https://orcid.org/0000-0002-0606-7930}{\includegraphics[scale=0.5]{figures/orcid.jpg}}}$, 
N. Thomas$^{13}$, 
S. Udry$^{4\,\href{https://orcid.org/0000-0001-7576-6236}{\includegraphics[scale=0.5]{figures/orcid.jpg}}}$, \newauthor
B. Ulmer$^{49}$, 
V. Van Grootel$^{30\,\href{https://orcid.org/0000-0003-2144-4316}{\includegraphics[scale=0.5]{figures/orcid.jpg}}}$, 
J. Venturini$^{4\,\href{https://orcid.org/0000-0001-9527-2903}{\includegraphics[scale=0.5]{figures/orcid.jpg}}}$, 
N.A. Walton$^{57\,\href{https://orcid.org/0000-0003-3983-8778}{\includegraphics[scale=0.5]{figures/orcid.jpg}}}$, 
E. Willett$^{2\,\href{https://orcid.org/0000-0002-7831-1402}{\includegraphics[scale=0.5]{figures/orcid.jpg}}}$, 
T.G. Wilson$^{25\,\href{https://orcid.org/0000-0001-8749-1962}{\includegraphics[scale=0.5]{figures/orcid.jpg}}}$
\\
(Affiliations listed after the references)
}
\date{Accepted. Received; in original form}
\pubyear{2023}
\begin{document}
\label{firstpage}
\pagerange{\pageref{firstpage}--\pageref{lastpage}}
\maketitle

\begin{abstract}
Observations of low-mass stars have frequently shown a disagreement between observed stellar radii and radii predicted by theoretical stellar structure models.
This ``radius inflation'' problem could have an impact on both stellar and exoplanetary science.
We present the final results of our observation programme with the \textit{CHEOPS} satellite to obtain high-precision light curves of eclipsing binaries with low mass stellar companions (EBLMs).
Combined with the spectroscopic orbits of the solar-type companion, we can derive the masses, radii and effective temperatures of 23 M-dwarf stars.
We use the \texttt{PYCHEOPS} data analysis software to analyse their primary and secondary occultations.
For all but one target, we also perform analyses with \textit{TESS} light curves for comparison.
We have assessed the impact of starspot-induced variation on our derived parameters and account for this in our radius and effective temperature uncertainties using simulated light curves.
We observe trends for inflation with both metallicity and orbital separation.
We also observe a strong trend in the difference between theoretical and observational effective temperatures with metallicity.
There is no such trend with orbital separation.
These results are not consistent with the idea that observed inflation in stellar radius combines with lower effective temperature to preserve the luminosity predicted by low-mass stellar models.
Our EBLM systems are high-quality and homogeneous measurements that can be used in further studies into radius inflation.
\end{abstract}




\section{Introduction}

\input{intro}

\section{Observations and Methods}
\label{sec:obs}

 Our targets are all detached eclipsing binary stars in which a solar-type star is eclipsed by an M-dwarf. 
 The observations were made as part of the \textit{CHEOPS} Guaranteed Time Observation (GTO) programme CH\_PR100037: Eclipsing binaries with very low mass stars between April 2020 and October 2022. 
 This programme sought to observe the primary and secondary eclipses of 23 EBLM systems, which was achieved for 21/23 targets.
 \textit{CHEOPS} observes targets from low-Earth orbit.
 Therefore observations are interrupted by the Earth occulting the target and by travelling through the South Atlantic Anomaly. 
 This results in gaps in the light curve of up to 44 and 19 minutes, respectively.
 The efficiency of each visit, i.e. the amount of observing time spent observing the object due to these gaps is shown in Table \ref{FullObsLog}.

    The raw data were processed using version 13.1 of the \textit{CHEOPS} data reduction pipeline \citep[DRP,][]{hoyer2020expected}.
    The DRP performs image correction for environmental and instrumental effects before performing aperture photometry of the target.
    As explained in \cite{hoyer2020expected}, the Gaia DR2 catalogue \citep{gaia2018gaia} is used by the DRP to simulate the field of view (FoV) of an observation in order to estimate the level of contamination present in the photometric aperture.
    The DRP also accounts for the rotating FoV of \textit{CHEOPS}, where other stars in the image can create ``smear'' trails and contaminate the photometric aperture.
    The smear effect is corrected by the DRP while the contamination produced by nearby stars is recorded in the DRP data products, allowing the user to include or ignore the contamination correction provided.
    The final photometry is extracted by the DRP using three different fixed aperture sizes labelled "RINF", "DEFAULT" and "RSUP" (at radii of 22.5, 25.0 and 30.0 pixels, respectively) and a further "OPTIMAL" aperture whose size is dependent upon the FoV contamination.
    The observed and processed data are made available on the Data Analysis Center for Exoplanets (DACE) web platform\footnote{The DACE platform is available at \url{http://dace.unige.ch}}.
    We downloaded our data from DACE using \texttt{PYCHEOPS}\footnote{\url{https://pypi.org/project/pycheops/}}, a \textsc{PYTHON} module developed for the analysis of data from the \textit{CHEOPS} mission \citep{pycheops}.
 The log of our observations including the aperture radius chosen to analyse each light curve are shown in Table \ref{FullObsLog}.
 We fitted the light curves from all four apertures and found that different choice of aperture radius has a negligible impact on the results.
 Therefore, we chose the aperture based on which gave the minimum mean absolute deviation (MAD) of the point-to-point differences in the light curve of the eclipse visit.

\begin{table*}

      \caption{A log of observation dates and details for each target visit. Effic. is the fraction of the observing interval covered by valid observations of the target. $R_{\rm ap}$ is the aperture radius used to compute the light curve analysed in this paper.}
         \label{FullObsLog}
    $$ 
         \begin{tabular}{lcrrrrr}
            \hline\hline
            \noalign{\smallskip}
            Event & \multicolumn{1}{c}{Target} & \multicolumn{1}{l}{Start Date (UTC)} & \multicolumn{1}{l}{Duration [s]} & \multicolumn{1}{l}{Effic. (\%)} & \multicolumn{1}{l}{File key} & \multicolumn{1}{l}{$R_{\rm ap}$ [pixels]}\\
            \noalign{\smallskip}
            \hline
            \noalign{\smallskip}
            Transit & J0057-19  & 2020-10-27T10:08:00 & 31586 & 78.9 & CH\_PR100037\_TG011401\_V0200 & 25.0 \\
            Eclipse &  & 2020-10-25T06:22:00 & 31824 & 82.1 & CH\_PR100037\_TG011301\_V0200 & 25.0 \\
            Transit & J0113+31  & 2020-11-24T15:37:00 & 49425 & 52.8 & CH\_PR100037\_TG011601\_V0200 & 40.0 \\
            Transit &  & 2021-10-19T00:16:00 & 49425 & 63.5 & CH\_PR100037\_TG017101\_V0200 & 40.0 \\
            Eclipse & & 2021-09-28T03:07:00 & 35379 & 57.9 & CH\_PR100037\_TG017201\_V0200 & 40.0 \\
            Transit & J0123+38 & 2020-10-17T16:16:00 & 45098 & 55.1 & CH\_PR100037\_TG011801\_V0200 & 22.5 \\
            Eclipse &  & 2020-11-14T13:10:00 & 45098 & 51.8 & CH\_PR100037\_TG011701\_V0200 & 22.5 \\
            Eclipse & & 2020-12-16T07:53:00 & 45098 & 54.7 & CH\_PR100037\_TG011702\_V0200 & 22.5 \\
            Transit & J0239-20 & 2020-11-01T15:40:00 & 30876 & 88.6 & CH\_PR100037\_TG012001\_V0200 & 25.0\\
            Eclipse &  & 2020-11-05T20:08:00 & 30224 & 95.0 & CH\_PR100037\_TG011901\_V0200 & 25.0\\
            Eclipse & & 2020-11-19T17:20:00 & 30224 & 74.0 & CH\_PR100037\_TG011902\_V0200 & 25.0 \\
            Transit & J0540-17 & 2020-12-07T08:36:00 & 37987 & 71.1 & CH\_PR100037\_TG012601\_V0200 & 18.0 \\
            Eclipse &  & 2020-12-04T08:10:00 & 38580 & 67.7 & CH\_PR100037\_TG012501\_V0200 & 18.0 \\
            Eclipse & & 2021-01-21T09:38:41 & 38580 & 55.7 & CH\_PR100037\_TG012502\_V0200 & 18.0 \\
            Eclipse & & 2021-01-27T09:19:41 & 38580 & 54.3 & CH\_PR100037\_TG012503\_V0200 & 18.0 \\
            Transit & J0546-18 & 2020-11-30T22:24:00 & 29927 & 69.4 & CH\_PR100037\_TG012801\_V0200 & 25.0 \\
            Eclipse &  & 2020-12-31T05:23:11 & 29987 & 66.8 & CH\_PR100037\_TG012701\_V0200 & 25.0 \\
            Eclipse & & 2021-01-09T19:36:00 & 29987 & 67.1 & CH\_PR100037\_TG012702\_V0200 & 25.0 \\
            Transit & J0719+25 & 2020-12-10T07:00:00 & 33483 & 55.0 & CH\_PR100037\_TG013001\_V0200 & 22.5 \\
            Eclipse$^{\dagger}$ &  & 2020-12-21T12:00:00 & 32713 & 62.2 & CH\_PR100037\_TG012901\_V0200 & 22.5 \\
            Eclipse & & 2021-02-03T20:51:00 & 33127 & 58.8 & CH\_PR100037\_TG017301\_V0200 & 22.5 \\
            Transit & J0941-31 & 2021-03-05T05:01:00 & 37217 & 74.0 & CH\_PR100037\_TG013401\_V0200 & 22.5 \\
            Eclipse &  & 2021-02-14T12:55:00 & 37512 & 91.3 & CH\_PR100037\_TG013301\_V0200 & 22.5 \\
            Transit & J0955-39 & 2021-04-12T14:56:00 & 30283 & 56.0 & CH\_PR100037\_TG013601\_V0200 & 22.5 \\
            Eclipse & & 2021-02-21T02:42:00 & 30224 & 69.0 & CH\_PR100037\_TG013501\_V0200 & 22.5 \\
            Transit & J1013+01 & 2021-01-29T15:13:00 & 28920 & 63.3 & CH\_PR100037\_TG013801\_V0200 & 30.0 \\
            Eclipse &  & 2021-03-18T09:41:00 & 28801 & 92.6 & CH\_PR100037\_TG013701\_V0200 & 30.0 \\
            Transit & J1305-31 & 2021-04-06T13:59:00 & 37098 & 90.5 & CH\_PR100037\_TG014001\_V0200 & 30.0 \\
            Eclipse &  & 2021-04-11T15:59:00 & 36387 & 90.7 & CH\_PR100037\_TG013901\_V0200 & 30.0 \\
            Eclipse & J1522+42 & 2021-04-16T02:33:43 & 34905 & 56.4 & CH\_PR100037\_TG016601\_V0200 & 25.0 \\
            Transit & J1559-05 & 2021-06-07T19:08:00 & 31705 & 92.7 & CH\_PR100037\_TG014401\_V0200 & 22.5 \\
            Eclipse & & 2020-04-18T08:17:00 & 31705 & 70.5 & CH\_PR100037\_TG014301\_V0200 & 22.5 \\
            Eclipse & & 2020-06-09T23:16:00 & 31705 & 95.5 & CH\_PR100037\_TG014302\_V0200 & 22.5 \\
            Eclipse & & 2022-06-01T21:13:00 & 31705  & 94.4 & CH\_PR100037\_TG014303\_V0200 & 22.5 \\
            Eclipse & & 2022-06-13T05:05:00 & 31705 & 76.9 & CH\_PR100037\_TG014304\_V0200 & 22.5 \\
            Transit & J1741+31 & 2020-06-13T08:20:00 & 27794 & 67.8 & CH\_PR100037\_TG014601\_V0200 & 30.0    \\
            Eclipse$^{\dagger}$ & & 2020-06-10T08:12:58 & 29098 & 63.0 & CH\_PR100037\_TG014501\_V0200 & 30.0  \\
            Transit & J1928-38 & 2021-06-09T16:14:00 & 45810 & 54.4 & CH\_PR100037\_TG014801\_V0200 & 22.5 \\
            Eclipse &  & 2021-06-20T12:20:00 & 47113 & 57.1 & CH\_PR100037\_TG014701\_V0200 & 22.5 \\
            Transit & J1934-42 & 2020-06-27T13:43:57 & 28387 & 60.7 & CH\_PR100037\_TG015001\_V0200 & 25.0 \\
            Eclipse & & 2020-07-13T09:47:00 & 28387 & 61.1 & CH\_PR100037\_TG014901\_V0200 & 25.0   \\
            Transit & J2040-41 & 2021-06-24T18:49:00 & 45395 & 52.8 & CH\_PR100037\_TG015201\_V0200 & 22.5 \\
            Eclipse &  & 2021-06-19T06:13:12 & 42609 & 53.0 & CH\_PR100037\_TG015101\_V0200 & 22.5 \\
            Eclipse & & 2021-09-13T22:40:00 & 42609 & 63.5 & CH\_PR100037\_TG015102\_V0200 & 22.5 \\
            Eclipse & J2046-40 & 2021-09-07T13:41:00 & 50195 & 60.8 & CH\_PR100037\_TG015301\_V0200 & 25.0 \\
            Eclipse &  & 2022-06-30T14:52:42 & 56906 & 57.8 & CH\_PR100037\_TG017401\_V0200 & 25.0 \\
            Transit & J2046+06 & 2020-08-28T22:08:00 & 35676 & 81.1 & CH\_PR100037\_TG015601\_V0200 & 25.0   \\
            Eclipse & & 2020-07-03T11:34:00 & 42313 & 66.7 & CH\_PR100037\_TG015501\_V0200 & 25.0  \\
            Eclipse & & 2021-07-22T13:59:00 & 42313 & 91.4 & CH\_PR100037\_TG015502\_V0200 & 25.0 \\
            Eclipse & & 2021-08-11T20:30:55 & 42313 & 94.0 & CH\_PR100037\_TG015503\_V0200 & 25.0 \\
            Transit & J2134+19 & 2020-07-16T15:22:00 & 46106 & 62.0 & CH\_PR100037\_TG015801\_V0200 & 22.5 \\
            Transit &  & 2020-09-21T08:50:00 & 47410 & 59.4 & CH\_PR100037\_TG017001\_V0200 & 22.5 \\
            Eclipse & & 2022-10-18T03:26:00 & 49163 & 56.8 & CH\_PR100037\_TG017501\_V0200 & 22.5 \\
            Transit & J2315+23 & 2021-09-27T12:04:00 & 41424 & 61.1 & CH\_PR100037\_TG016001\_V0200 & 22.5 \\
            Eclipse &  & 2021-09-13T01:29:00 & 39172 & 71.3 & CH\_PR100037\_TG016801\_V0200 & 22.5 \\
            Transit & J2343+29 & 2021-09-17T21:03:59 & 33483 & 71.5 & CH\_PR100037\_TG016201\_V0200 & 25.0 \\
            Eclipse &  & 2021-09-09T17:47:00 & 36979 & 67.1 & CH\_PR100037\_TG016101\_V0200 & 25.0 \\
            Transit & J2359+44 & 2020-11-28T12:20:00 & 60507 & 53.3 & CH\_PR100037\_TG016401\_V0200 & 26.0 \\
            Eclipse &  & 2020-11-11T08:37:00 & 33483 & 60.1 & CH\_PR100037\_TG016301\_V0200 & 26.0 \\
            \noalign{\smallskip}
            \hline
            \multicolumn{7}{@{}l}{$^{\dagger}$ Does not cover the phase of superior conjunction.}
         \end{tabular}
    $$
\end{table*}
    
    The \textit{TESS} survey is split into overlapping $90^{\circ} \times 24^{\circ}$ degree sky sectors over the northern and southern hemispheres with one sector being observed for approximately one month.
    We used 2-minute cadence data observed as part of the EBLM group's \textit{TESS} Guest Investigator (GI) programmes G011278, G022253, G03216, G04157 and G05024.
    Our targets were also observed by other groups under GI programmes G022039, G022062, G022156, G03251, G03272, G04171, G04191, G04234, G05003 and G05112.
    Data were reduced by the Science Processing Operations Center Pipeline \citep[SPOC;][]{jenkins2016tess} and made available from the Mikulski Archive for Space Telescopes (MAST)\footnote{\url{https://mast.stsci.edu}} web service.
    We used the Pre-search Data Conditioned Simple Aperture Photometry (PDCSAP) flux data in our analysis.
    Any cadences in the light curve with severe quality issues were ignored using the ``default'' bitmask 175 \citep{tenenbaum}.
    We only used segments of the \textit{TESS} light curve within one eclipse\footnote{Here referring to the primary eclipse of the host star by the secondary in transit or the secondary eclipse due to the occultation of the secondary star.} duration of the time of mid-eclipse of each occultation. 
    To remove trends in the light curve, the segments were divided by a linear polynomial model fitted to the data on either side of the masked-out eclipse. 
    The out-of-eclipse data would be used to quantify the effect flux variation had on our fitted results.

    \subsection{Derivation of primary effective temperature and metallicity}
    The spectroscopic stellar parameters ($T_{\mathrm{eff}}$, $\log g$, microturbulence ($\xi_{\mathrm{t}}$), [Fe/H]) and their respective uncertainties were estimated by using ARES+MOOG, following the same methodology as described in \citet[][]{sousa21, sousa14, santos2013}. 
    We used co-added spectra from individual observations performed with the SOPHIE spectrograph for EBLM~J0719+25, EBLM~J1522+42, EBLM~J1741+31, EBLM~J2134+19, EBLM~J2315+23, and EBLM~J2359+44, co-added CORALIE\footnote{Available from the ESO science archive facility \url{http://archive.eso.org/}} spectra for EBLM~J0540-17 and EBLM~J2046-40 obtained from \citet{triaud2017}, co-added FIES spectra \citep{Telting} for EBLM~J0113+31 and EBLM~J0123+38 and with HARPS observations from ESO programme 1101.C-0721 for EBLM~J0941-31, EBLM~1305-31, EBLM~1928-38, EBLM~J1934$-$42, EBLM~J2040-41 and EBLM~J2046+06. 
    We used the ARES code\footnote{The last version of ARES code (ARES v2) can be downloaded at \url{https://github.com/sousasag/ARES}} \citep{Sousa-07, sousa15} to measure equivalent widths (EW) of iron lines from the list of lines presented in \citet[][]{Sousa-08}. For EBLM~J2343$+$29 we instead used the appropriate list of lines presented in \citet[][]{tsantaki13} as the star has a lower temperature ($T_{\mathrm{eff}} < 5200 K$). A minimization process assuming ionization and excitation equilibrium is used to find convergence for the best set of spectroscopic parameters. In this process we use a grid of Kurucz model atmospheres \citep{Kurucz-93} and the radiative transfer code MOOG \citep{Sneden-73}.
    There were some targets we analysed with different methods due to difficulties with the spectra e.g. low signal-to-noise ratio (SNR).
    Using SALT spectra for EBLM~J0057-19, EBLM~J0239-20 and using CORALIE spectra for EBLM~J1559-05, we modeled the stellar fundamental parameters using the software \texttt{SME}\footnote{\url{http://www.stsci.edu/~valenti/sme.html}} \citep[Spectroscopy Made Easy;][]{vp96, pv2017}. 
    \texttt{SME} computes synthetic spectra with atomic and molecular line data from \texttt{VALD}\footnote{\url{http://vald.astro.uu.se}}  \citep{Ryabchikova2015} which is compared to the observations. 
    We chose the stellar atmosphere grid Atlas12 \citep{Kurucz2013} and modelled  $\rm T_{eff}$, $\rm log\,g_{1}$, abundances and $\rm v\,sin\,i$ one parameter at a time.
    For EBLM J0546-18, EBLM~J0955-39 and EBLM~J1031+01 we used co-added CORALIE spectra and applied a wavelet decomposition method where we compare the coefficients from a wavelet decomposition to those from a grid of model spectra. Those model spectra were synthesised using the code \texttt{SPECTRUM} \citep{1994AJ....107..742G}, MARCS model atmospheres \citep{2008A&A...486..951G}, as well as the atomic line list version 5 of the Gaia ESO survey \citep{2015PhyS...90e4010H}. The method is detailed in \cite{gill18} and has been found to deliver robust measurements of effective temperature and metallicity for spectra with relatively low SNR ($\rm SNR \gtrapprox 40$). 

\begin{table*}

      \caption{The primary stellar parameters used in deriving our final results. Sp. Type is the estimated spectral type of the primary star. The primary effective temperature and metallicity were derived by the \textit{CHEOPS} TS3 - Target Characterisation working group using fitted spectra. The radial velocity semi-amplitude ($K_1$), eccentricity (e) and argument of periastron ($\omega$) values were obtained from \protect\citet{triaud2017}, \protect\citet{martin2019}, RV fits of individual targets or from our own \textit{ELLC} fits of radial velocity data.}
     \label{Primparams}
    $ 
         \begin{array}{lcrrrrrrr}
            \hline\hline
            \noalign{\smallskip}
            \multicolumn{1}{c}{\text{Target}} & \multicolumn{1}{c}{\text{Sp. Type}} & \multicolumn{1}{l}{{\rm V}} & \multicolumn{1}{l}{{\rm T}_{\rm eff,1}} & \multicolumn{1}{l}{\text{[Fe/H]}} & \multicolumn{1}{l}{K_1} & \multicolumn{1}{l}{P} & \multicolumn{1}{l}{e} & \multicolumn{1}{l}{\omega} \\
            & &\multicolumn{1}{l}{[{\rm mag.}]} &\multicolumn{1}{l}{[K]} & \multicolumn{1}{l}{[{\rm dex.}]} & \multicolumn{1}{l}{[{\rm km/s}]} & \multicolumn{1}{l}{[\text{days}]} & & \multicolumn{1}{l}{[{\rm deg}]}  \\
            \noalign{\smallskip}
            \hline
            \noalign{\smallskip}
            \text{J0057-19} & \text{G6V} & 11.65 & 5580 \pm 150 & 0.23 \pm 0.09 & 15.523 \pm 0.025 & 4.30055 \pm 0.00015 & 0.0 & --  \\
            \text{J0113+31} & \text{F9V} & 10.11 & 6025 \pm 76 & -0.31 \pm 0.05 & 15.861 \pm 0.010 & 14.276843 \pm 0.000003 & 0.3088 \pm 0.0005 & 279.00 \pm 0.03 \\
            \text{J0123+38} & \text{F8V} & 12.10 & 6182 \pm 91 & 0.452 \pm 0.070 & 27.59 \pm 0.17 & 7.952938 \pm 0.000006 & 0.0 & -- \\
            \text{J0239-20} & \text{G2V} & 10.63 & 5758 \pm 100  & 0.27 \pm 0.12 & 21.316 \pm 0.036 & 2.778691 \pm 0.000001 & 0.0 & -- \\
            \text{J0540-17} & \text{F7V} & 11.31 & 6290 \pm 77  & -0.04 \pm 0.05 & 16.199 \pm 0.010 & 6.004940 \pm 0.000003 & 0.0 & -- \\
            \text{J0546-18} & \text{F8V} & 12.15 & 6180 \pm 80 & -0.45 \pm 0.08 & 26.15 \pm 0.10 & 3.191919 \pm 0.000034 & 0.0 & -- \\
            \text{J0719+25} & \text{F9V} & 10.96 & 6026 \pm 67 & 0.04 \pm 0.05  & 15.02 \pm 0.04 & 7.456295 \pm 0.000045 & 0.0730 \pm 0.0045  & -155.8 \pm 5.4 \\
            \text{J0941-31} & \text{F5V} & 11.08 & 6504 \pm 101 & 0.078 \pm 0.069 & 21.312 \pm 0.036 & 5.54563 \pm 0.000018 & 0.2006 \pm 0.0017 & 5.02 \pm 0.52 \\
            \text{J0955-39} & \text{F6V} & 12.90 & 6340 \pm 80 & -0.24 \pm 0.08 & 21.446 \pm 0.034 & 5.3136 \pm 0.000012 & 0.0 & -- \\
            \text{J1013+01} & \text{K1V} & 11.87 & 5200 \pm 80 & 0.09 \pm 0.08 & 23.193 \pm 0.080 & 2.892273 \pm 0.0000024 & 0.0 & --  \\
            \text{J1305-31} & \text{G0V} & 12.10 & 5913 \pm 64 & 0.201 \pm 0.044 & 22.402 \pm 0.011 & 10.61913 \pm 0.000015 & 0.0374 \pm 0.0005 & -153.52 \pm 0.79  \\
            \text{J1522+42} & \text{G2V} & 11.66 & 5738 \pm 64 & -0.061 \pm 0.044  & 16.31 \pm 0.23 & 7.661343 \pm 0.000003 & 0.1386 \pm 0.0067 & -130.28 \pm 3.03 \\
            \text{J1559-05} & \text{F8V} & 9.69 & 6204 \pm 100 & 0.19 \pm 0.09  & 18.063 \pm 0.042 & 3.760075 \pm 0.0000023 & 0.0 & --  \\
            \text{J1741+31} & \text{F6V} & 11.70 & 6376 \pm 72 & 0.09 \pm 0.05 & 37.140 \pm 0.040 & 7.71263 \pm 0.00004 & 0.3009 \pm 0.0009 & 56.81 \pm 0.19  \\
            \text{J1928-38} & \text{G4V} & 11.20 & 5687 \pm 62 & -0.009 \pm 0.042 & 17.269 \pm 0.005 & 23.32286 \pm 0.000071 & 0.0735 \pm 0.0002 & -137.24 \pm 0.19 \\
            \text{J1934-42} & \text{G5V} & 12.62 & 5648 \pm 68 & 0.288 \pm 0.046 & 18.621 \pm 0.009 & 6.35251 \pm 0.00001 & 0.0 & -- \\
            \text{J2040-41} & \text{G2V} & 11.49 & 5790 \pm 63 & -0.206 \pm 0.043 & 12.462 \pm 0.004 & 14.45626 \pm 0.000031 & 0.2265 \pm 0.0003 & -36.82 \pm 0.10 \\
            \text{J2046-40} & \text{G2V} & 11.49 & 5763 \pm 75 & 0.337 \pm 0.054 & 11.986 \pm 0.012 & 37.013621 \pm 0.000023 & 0.4732 \pm 0.0005 & 155.77 \pm 0.06 \\
            \text{J2046+06} & \text{F7V} & 9.86 & 6302 \pm 70 & 0.000 \pm 0.048 & 15.548 \pm 0.006 & 10.10779 \pm 0.00001 & 0.3436 \pm 0.0003 & 108.92 \pm 0.08 \\
            \text{J2134+19} & \text{G8V} & 11.85 & 5421 \pm 64 & -0.57 \pm 0.05 & 26.706 \pm 0.892 & 16.58558 \pm 0.00005 & 0.2512 \pm 0.0270 & 35.18\pm 3.65\\
            \text{J2315+23} & \text{F9V} & 11.56 & 6027 \pm 66 & 0.02 \pm 0.05 & 19.98 \pm 0.46 & 9.13105 \pm 0.000119 & 0.149 \pm 0.001 & 147.23 \pm 0.34 \\
            \text{J2343+29} & \text{K2V} & 10.59 & 4984 \pm 87 & 0.11 \pm 0.05 & 8.418 \pm 0.003 & 16.95353 \pm 0.00005 & 0.1604 \pm 0.0003 & 78.41 \pm 0.09 \\
            \text{J2359+44} & \text{F2V} & 10.59 & 6799 \pm 83 & 0.12 \pm 0.05 & 23.62 \pm 0.08 & 11.3627 \pm 0.0027 & 0.4773 \pm 0.0010  & -94.29 \pm 0.06 \\
            \noalign{\smallskip}
            \hline
         \end{array}
    $
\end{table*}

\subsection{Complementary observations using SOPHIE}
    The semi-amplitude of the primary star's spectroscopic orbit, $K_1$, is required to estimate the secondary star's mass.
    For 14 out of 23 targets we used values of $K_1$ that are presented in the first paper of the `Binaries Escorted By Orbiting Planets' series \citet[][BEBOP I]{martin2019} or in \citet[][EBLM IV]{triaud2017}. 
    For EBLM~J0113+31 we used values published by \citet{maxted21}.
    For EBLM~J2343+29 we derived values using a simultaneous fit of NITES photometry \citep{mccormac} and radial velocities (RVs) using the \textsc{PYTHON} module \texttt{ELLC} \citep{maxted2016ellc}.
    The radial velocity (RV) data derives from an analysis of PARAS and SOPHIE data by \citet{chaturvedi} and of FIES data observed on opticon proposals 2011B$\_$017 and 2012A$\_$002.
    For the 7 remaining systems (J0123+38, J0719+25, J1522+42, J1741+31, J2134+19, J2315+23 and J2359+44), we used as-of-yet unpublished RV measurements obtained with the SOPHIE high-resolution \'echelle spectrograph \citep{Perruchot08} mounted on the 193cm telescope at the Observatoire de Haute-Provence (France). These were obtained in the context of BEBOP, a radial-velocity survey for circumbinary planets orbiting single-lined eclipsing binaries. All observations were performed with one fibre on the science target and one fibre on the sky to remove background contamination originating from the Moon. All science and sky spectra were reduced using the SOPHIE Data Reduction Software (DRS) and cross-correlated with a mask to obtain radial-velocities. We used a G2 mask for G and F dwarfs, and a K5 mask for K dwarfs. These methods are described in \citet{Baranne96}, \citet{Courcol15} and Chapter 2 of \citet{heidari2022}, and have been shown to produce precisions and accuracies of a few meters per seconds across F, G and K spectral types \citep[e.g.][]{Bouchy2013,Hara20}, well below what we typically obtained on this system.
    All radial velocity observations are high-resolution and stable at the 1${\rm m s}^{\rm -1}$ level and so should be consistent measuring the 1${\rm km s}^{\rm -1}$ amplitudes in this study.
    We used {\fontfamily{qcr}\selectfont ELLC} to model the radial velocity.
    In our fit of the Keplerian orbit we accounted for jitter by applying a weight in our log-likelihood function.
    All of the primary stars' stellar and orbital parameters are listed in Table \ref{Primparams}.

\section{Analysis}
\label{sec:ANA}

\input{ana}

\section{Simulations of starspot activity and derived uncertainties}
\label{sec:spt}

\input{spt}

\input{res}

\section*{Acknowledgements}

CHEOPS is an ESA mission in partnership with Switzerland with important contributions to the payload and the ground segment from Austria, Belgium, France, Germany, Hungary, Italy, Portugal, Spain, Sweden, and the United Kingdom. The CHEOPS Consortium would like to gratefully acknowledge the support received by all the agencies, offices, universities, and industries involved. Their flexibility and willingness to explore new approaches were essential to the success of this mission.
Funding for the TESS mission is provided by NASA’s Science Mission directorate.
Based on observations collected at the European Southern Observatory under ESO programmes 1101.C-0721, 60.A-9022(A) and 072.C-0002(D).
Based on observations made with the Nordic Optical Telescope, owned in collaboration by the University of Turku and Aarhus University, and operated jointly by Aarhus University, the University of Turku and the University of Oslo, representing Denmark, Finland and Norway, the University of Iceland and Stockholm University at the Observatorio del Roque de los Muchachos, La Palma, Spain, of the Instituto de Astrofisica de Canarias.
Some of the observations reported in this paper were obtained with the Southern African Large Telescope (SALT).
This publication makes use of The Data \& Analysis Center for Exoplanets (DACE), which is a facility based at the University of Geneva (CH) dedicated to extrasolar planets data visualisation, exchange and analysis. 
DACE is a platform of the Swiss National Centre of Competence in Research (NCCR) PlanetS, federating the Swiss expertise in Exoplanet research. 
The DACE platform is available at \url{https://dace.unige.ch}.

MIS and PFLM acknowledges support from STFC research grant numbers ST/M001040/1 and ST/T506175/1. 
This research is supported from the European Research Council (ERC) under the European Union's Horizon 2020 research and innovation programme (grant agreement n$^\circ$ 803193/BEBOP), by a Leverhulme Trust Research Project Grant (n$^\circ$ RPG-2018-418), and by observations obtained at the Observatoire de Haute-Provence (CNRS), France (PI Santerne). 
S.G.S. acknowledge support from FCT through FCT contract nr. CEECIND/00826/2018 and POPH/FSE (EC). 
This project has received funding from the European Research Council (ERC) under the European Union’s Horizon 2020 research and innovation programme (project {\sc Four Aces}. 
grant agreement No 724427). It has also been carried out in the frame of the National Centre for Competence in Research PlanetS supported by the Swiss National Science Foundation (SNSF). DE acknowledges financial support from the Swiss National Science Foundation for project 200021\_200726. 
SH gratefully acknowledges CNES funding through the grant 837319. 
The French group acknowledges financial support from the French Programme National de Plan\'etologie (PNP, INSU). 
Support for DVM was provided by NASA through the NASA Hubble Fellowship grant HF2-51464 awarded by the Space Telescope Science Institute, which is operated by the Association of Universities for Research in Astronomy, Inc., for NASA, under contract NAS5-26555. 
MRS acknowledges support from the UK Science and Technology Facilities Council (ST/T000295/1). 
YAl acknowledges support from the Swiss National Science Foundation (SNSF) under grant 200020\_192038. 
RAl, DBa, EPa, and IRi acknowledge financial support from the Agencia Estatal de Investigación of the Ministerio de Ciencia e Innovación MCIN/AEI/10.13039/501100011033 and the ERDF “A way of making Europe” through projects PID2019-107061GB-C61, PID2019-107061GB-C66, PID2021-125627OB-C31, and PID2021-125627OB-C32, from the Centre of Excellence “Severo Ochoa'' award to the Instituto de Astrofísica de Canarias (CEX2019-000920-S), from the Centre of Excellence “María de Maeztu” award to the Institut de Ciències de l’Espai (CEX2020-001058-M), and from the Generalitat de Catalunya/CERCA programme. 
S.C.C.B. acknowledges support from FCT through FCT contracts nr. IF/01312/2014/CP1215/CT0004. 
XB, SC, DG, MF and JL acknowledge their role as ESA-appointed CHEOPS science team members. 
LBo, VNa, IPa, GPi, RRa, and GSc acknowledge support from CHEOPS ASI-INAF agreement n. 2019-29-HH.0. 
This work has been carried out within the framework of the NCCR PlanetS supported by the Swiss National Science Foundation under grants 51NF40$\_$182901 and 51NF40$\_$205606. This project has received funding from the European Research Council (ERC) under the European Union's Horizon 2020 research and innovation programme (project {\sc Spice Dune}, grant agreement No 947634). 
ABr was supported by the SNSA. 
A.C, X.D., T.F, acknowledge funding from the French ANR under contract number ANR\-18\-CE31\-0019 (SPlaSH). This work is supported by the French National Research Agency in the framework of the Investissements d'Avenir program (ANR-15-IDEX-02), through the funding of the ``Origin of Life'' project of the Grenoble-Alpes University. 
ACC and TWi acknowledge support from STFC consolidated grant numbers ST/R000824/1 and ST/V000861/1, and UKSA grant number ST/R003203/1. 
P.E.C. is funded by the Austrian Science Fund (FWF) Erwin Schroedinger Fellowship, program J4595-N.
This project was supported by the CNES. 
The Belgian participation to CHEOPS has been supported by the Belgian Federal Science Policy Office (BELSPO) in the framework of the PRODEX Program, and by the University of Liège through an ARC grant for Concerted Research Actions financed by the Wallonia-Brussels Federation. 
L.D. is an F.R.S.-FNRS Postdoctoral Researcher. 
This work was supported by FCT - Fundação para a Ciência e a Tecnologia through national funds and by FEDER through COMPETE2020 - Programa Operacional Competitividade e Internacionalizacão by these grants: UID/FIS/04434/2019, UIDB/04434/2020, UIDP/04434/2020, PTDC/FIS-AST/32113/2017 \& POCI-01-0145-FEDER- 032113, PTDC/FIS-AST/28953/2017 \& POCI-01-0145-FEDER-028953, PTDC/FIS-AST/28987/2017 \& POCI-01-0145-FEDER-028987, O.D.S.D. is supported in the form of work contract (DL 57/2016/CP1364/CT0004) funded by national funds through FCT. 
B.-O. D. acknowledges support from the Swiss State Secretariat for Education, Research and Innovation (SERI) under contract number MB22.00046. 
MF and CMP gratefully acknowledge the support of the Swedish National Space Agency (DNR 65/19, 174/18). 
DG gratefully acknowledges financial support from the CRT foundation under Grant No. 2018.2323 ``Gaseousor rocky? Unveiling the nature of small worlds''. 
M.G. is an F.R.S.-FNRS Senior Research Associate. 
MNG is the ESA CHEOPS Project Scientist and Mission Representative, and as such also responsible for the Guest Observers (GO) Programme. MNG does not relay proprietary information between the GO and Guaranteed Time Observation (GTO) Programmes, and does not decide on the definition and target selection of the GTO Programme. 
CHe acknowledges support from the European Union H2020-MSCA-ITN-2019 under Grant Agreement no. 860470
(CHAMELEON). 
K.W.F.L. was supported by Deutsche Forschungsgemeinschaft grants RA714/14-1 within the DFG Schwerpunkt SPP 1992, Exploring the Diversity of Extrasolar Planets. 
This work was granted access to the HPC resources of MesoPSL financed by the Region Ile de France and the project Equip@Meso (reference ANR-10-EQPX-29-01) of the programme Investissements d'Avenir supervised by the Agence Nationale pour la Recherche. 
ML acknowledges support of the Swiss National Science Foundation under grant number PCEFP2\_194576. 
E.M. acknowledges funding from FAPEMIG under project number APQ-02493-22 and research productivity grant number 309829/2022-4 awarded by the CNPq, Brazil. 
This work was also partially supported by a grant from the Simons Foundation (PI Queloz, grant number 327127). 
NCSa acknowledges funding by the European Union (ERC, FIERCE, 101052347). Views and opinions expressed are however those of the author(s) only and do not necessarily reflect those of the European Union or the European Research Council. Neither the European Union nor the granting authority can be held responsible for them. 
GyMSz acknowledges the support of the Hungarian National Research, Development and Innovation Office (NKFIH) grant K-125015, a PRODEX Experiment Agreement No. 4000137122, the Lend\"ulet LP2018-7/2021 grant of the Hungarian Academy of Science and the support of the city of Szombathely. 
V.V.G. is an F.R.S-FNRS Research Associate. 
NAW acknowledges UKSA grant ST/R004838/1.

Many thanks to PFLM and DM for their proofreading of this paper.
We also thank the anonymous referee for their many suggestions and questions of this paper, through which it has doubtlessly been improved.

\section*{Data Availability}

The CHEOPS data underlying this article are available in its online supplementary material and are publicly available via the \href{http://dace.unige.ch/}{Data Analysis Center for
Exoplanets} web platform, as well as from the VizieR data base of astronomical catalogues at the Centre de Données astronomiques de Strasbourg (\href{http://cds.u-strasbg.fr/}). 

This paper includes data collected by the TESS mission, which is publicly available from the Mikulski Archive for Space Telescopes (MAST) at the Space Telescope Science Institure (STScI) (\url{https://mast.stsci.edu}). 
Funding for the TESS mission is provided by the NASA Explorer Program directorate. 
STScI is operated by the Association of Universities for Research in Astronomy, Inc., under NASA contract NAS 5–26555.
We acknowledge the use of public TESS Alert data from pipelines at the TESS Science Office and at the TESS Science Processing Operations Center.

SOPHIE high-resolution spectra are available through the data archives of the Observatoire de Haute-Provence via \url{http://atlas.obs-hp.fr/}. Programme ID were 18B.PNP.SAN1, and 19A.PNP.SANT. 



\bibliographystyle{mnras}
\bibliography{example} 

\bigskip

\noindent
\hrulefill \\
$^{1}$ Astrophysics Group, Lennard Jones Building, Keele University, Staffordshire, ST5 5BG, United Kingdom \\
$^{2}$ School of Physics and Astronomy, University of Birmingham, Edgbaston, Birmingham B15 2TT, UK \\
$^{3}$ Instituto de Astrofisica e Ciencias do Espaco, Universidade do Porto, CAUP, Rua das Estrelas, 4150-762 Porto, Portugal \\
$^{4}$ Observatoire Astronomique de l'Université de Genève, Chemin Pegasi 51, CH-1290 Versoix, Switzerland \\
$^{5}$ Centre Vie dans l’Univers, Faculté des sciences, Université de Genève, Quai Ernest-Ansermet 30, 1211 Genève 4, Switzerland \\
$^{6}$ Aix Marseille Univ, CNRS, CNES, LAM, 38 rue Frédéric Joliot-Curie, 13388 Marseille, France \\
$^{7}$ Department of Astronomy, Stockholm University, AlbaNova University Center, 10691 Stockholm, Sweden \\
$^{8}$ Department of Physics \& Astronomy, Tufts University, Medford, MA 02155, USA \\
$^{9}$ Department of Physics, University of Warwick, Gibbet Hill Road, Coventry CV4 7AL, United Kingdom \\
$^{10}$ Department of Space, Earth and Environment, Chalmers University of Technology, Onsala Space Observatory, 439 92 Onsala, Sweden \\
$^{11}$ School of Physical Sciences, The Open University, Milton Keynes, MK7 6AA, UK \\
$^{12}$ Center for Space and Habitability, University of Bern, Gesellschaftsstrasse 6, 3012 Bern, Switzerland \\
$^{13}$ Weltraumforschung und Planetologie, Physikalisches Institut, University of Bern, Gesellschaftsstrasse 6, 3012 Bern, Switzerland \\
$^{14}$ Instituto de Astrofisica de Canarias, Via Lactea s/n, 38200 La Laguna, Tenerife, Spain \\
$^{15}$ Departamento de Astrofisica, Universidad de La Laguna, Astrofísico Francisco Sanchez s/n, 38206 La Laguna, Tenerife, Spain \\
$^{16}$ Institut de Ciencies de l'Espai (ICE, CSIC), Campus UAB, Can Magrans s/n, 08193 Bellaterra, Spain \\
$^{17}$ Institut d’Estudis Espacials de Catalunya (IEEC), Gran Capità 2-4, 08034 Barcelona, Spain \\
$^{18}$ Admatis, 5. Kandó Kálmán Street, 3534 Miskolc, Hungary \\
$^{19}$ Depto. de Astrofisica, Centro de Astrobiologia (CSIC-INTA), ESAC campus, 28692 Villanueva de la Cañada (Madrid), Spain \\
$^{20}$ Departamento de Fisica e Astronomia, Faculdade de Ciencias, Universidade do Porto, Rua do Campo Alegre, 4169-007 Porto, Portugal \\
$^{21}$ Space Research Institute, Austrian Academy of Sciences, Schmiedlstrasse 6, A-8042 Graz, Austria \\
$^{22}$ Université Grenoble Alpes, CNRS, IPAG, 38000 Grenoble, France \\
$^{23}$ INAF, Osservatorio Astronomico di Padova, Vicolo dell'Osservatorio 5, 35122 Padova, Italy \\
$^{24}$ Université de Paris Cité, Institut de physique du globe de Paris, CNRS, 1 Rue Jussieu, F-75005 Paris, France \\
$^{25}$ Centre for Exoplanet Science, SUPA School of Physics and Astronomy, University of St Andrews, North Haugh, St Andrews KY16 9SS, UK \\
$^{26}$ Institute of Planetary Research, German Aerospace Center (DLR), Rutherfordstrasse 2, 12489 Berlin, Germany \\
$^{27}$ INAF, Osservatorio Astrofisico di Torino, Via Osservatorio, 20, I-10025 Pino Torinese To, Italy \\
$^{28}$ Centre for Mathematical Sciences, Lund University, Box 118, 221 00 Lund, Sweden \\
$^{29}$ Astrobiology Research Unit, Université de Liège, Allée du 6 Août 19C, B-4000 Liège, Belgium \\
$^{30}$ Space sciences, Technologies and Astrophysics Research (STAR) Institute, Université de Liège, Allée du 6 Août 19C, 4000 Liège, Belgium \\
$^{31}$ Leiden Observatory, University of Leiden, PO Box 9513, 2300 RA Leiden, The Netherlands \\
$^{32}$ Dipartimento di Fisica, Universita degli Studi di Torino, via Pietro Giuria 1, I-10125, Torino, Italy \\
$^{33}$ Department of Astrophysics, University of Vienna, Türkenschanzstrasse 17, 1180 Vienna, Austria \\
$^{34}$ ESTEC, European Space Agency, Keplerlaan 1, 2201AZ, Noordwijk, NL \\
$^{35}$ Institut d'astrophysique de Paris, UMR7095 CNRS, Université Pierre \& Marie Curie, 98bis blvd. Arago, 75014 Paris, France \\
$^{36}$ Observatoire de Haute-Provence, CNRS, Universit\'e d'Aix-Marseille, 04870 Saint-Michel l'Observatoire, France \\
$^{37}$ Laboratoire J.-L. Lagrange, Observatoire de la C\^ote d'Azur (OCA), Universite de Nice-Sophia Antipolis (UNS), CNRS, Campus Valrose, 06108 Nice Cedex 2, France. \\
$^{38}$ Institute for Theoretical Physics and Computational Physics, Graz University of Technology, Petersgasse 16, 8010 Graz, Austria \\
$^{39}$ Sorbonne Université, CNRS, UMR 7095, Institut d’Astrophysique de Paris, 98 bis bd Arago, 75014 Paris, France \\
$^{40}$ LESIA, Observatoire de Paris, Université PSL, CNRS, Sorbonne Université, Université Paris Cité, 5 place Jules Janssen, 92195 Meudon, France \\
$^{41}$ Konkoly Observatory, Research Centre for Astronomy and Earth Sciences, 1121 Budapest, Konkoly Thege Miklós út 15-17, Hungary \\
$^{42}$ ELTE E\"otv\"os Lor\'and University, Institute of Physics, P\'azm\'any P\'eter s\'et\'any 1/A, 1117 Budapest, Hungary \\
$^{43}$ Centre for Exoplanets and Habitability, University of Warwick, Gibbet Hill Road, Coventry CV4 7AL, UK \\
$^{44}$ IMCCE, UMR8028 CNRS, Observatoire de Paris, PSL Univ., Sorbonne Univ., 77 av. Denfert-Rochereau, 75014 Paris, France \\
$^{45}$ Laborat\'{o}rio Nacional de Astrof\'{i}sica, Rua Estados Unidos 154, 37504-364, Itajub\'{a} - MG, Brazil \\
$^{46}$ Centrum Astronomiczne im. Mikołaja Kopernika, Polish Academy of Sciences, Bartycka 18, 00-716, Warsaw, Poland \\
$^{47}$ Univ. de Toulouse, CNRS, IRAP, 14 av. Belin, 31400 Toulouse,France \\
$^{48}$ INAF, Osservatorio Astrofisico di Catania, Via S. Sofia 78, 95123 Catania, Italy \\
$^{49}$ Institute of Optical Sensor Systems, German Aerospace Center (DLR), Rutherfordstrasse 2, 12489 Berlin, Germany \\
$^{50}$ Dipartimento di Fisica e Astronomia "Galileo Galilei", Universita degli Studi di Padova, Vicolo dell'Osservatorio 3, 35122 Padova, Italy \\
$^{51}$ ETH Zurich, Department of Physics, Wolfgang-Pauli-Strasse 2, CH-8093 Zurich, Switzerland \\
$^{52}$ Cavendish Laboratory, JJ Thomson Avenue, Cambridge CB3 0HE, UK \\
$^{53}$ Zentrum für Astronomie und Astrophysik, Technische Universität Berlin, Hardenbergstr. 36, D-10623 Berlin, Germany \\
$^{54}$ Institut fuer Geologische Wissenschaften, Freie Universitaet Berlin, Maltheserstrasse 74-100,12249 Berlin, Germany \\
$^{55}$ ELTE E\"otv\"os Lor\'and University, Gothard Astrophysical Observatory, 9700 Szombathely, Szent Imre h. u. 112, Hungary \\
$^{56}$ HUN-REN-ELTE Exoplanet Research Group, 9700 Szombathely, Szent Imre h. u. 112, Hungary \\
$^{57}$ Institute of Astronomy, University of Cambridge, Madingley Road, Cambridge, CB3 0HA, United Kingdom



\bsp	
\label{lastpage}
\end{document}

%% file: intro.tex
In exoplanet observation, the correct characterisation of the stellar host is of great importance.
The properties of the exoplanet such as mass and radius are most commonly inferred from their impact upon their host star, as seen with the transit and radial velocity methods \citep{2009MNRAS.394..272S}.
Increased accuracy of the mass and size of the host leads to increased accuracy in deriving the masses and sizes of any orbiting bodies.
The properties of the host star are most commonly derived by finding the best fit between observed properties and stellar evolution models \citep[e.g.][]{Baraffe98,Dotter}.
Therefore, any uncertainty in the models would give rise to systematic errors in inferred stellar properties and thus those of the exoplanet.
This has become a potential issue regarding low-mass star systems' recent popularity as targets for exoplanet observation \citep{charbonneau2007dynamics, quirrenbach2014carmenes,gillon2017seven,delrez}.
Upon observing more and more low mass stars a concerning issue has been identified.
A significant fraction of the stellar population at low masses have been observed with a radii that differs significantly from those predicted by theoretical stellar models. 

First observed in the 1970s \citep{hoxie70,hoxie73,lacy1977}, this finding has continued to be observed ever since \citep{Popper,Clausen,torres2002,casagrande2008,torres2010,kraus2011,birkby2012,feiden2012,nefs2013,spada2013,torres2013,chen2014,dittmann2017,kesseli2018,swayne21,2022Galax..10...98M,Jennings}, being termed the ``radius inflation'' problem.
Along with claims of radius inflation are reports of effective temperatures that are too cool compared to stellar models, a trend clearly visible in the mass-effective temperature diagram  displayed in \citet{parsons2018scatter}.
The underprediction of effective temperature when combined with the overprediction of radius was suggested by \citet{hoxie70} to balance to give constant luminosity. 
This hypothesis, that luminosities are being predicted accurately by stellar models for low-mass stars has been explored in many studies since \citep{Delfosse00,torres2002,ribas2006,Torres2006,Torres2007}.
However, disagreements between theoretical and observed mass-luminosity relations \citep{Mann} suggest that this balance is only accurate to a few percent.

Multiple theories have been put forth to explain radius inflation.
One of these has has been stellar activity.
It has been proposed that sizable magnetic activity could inhibit convection \citep{feiden2013} transferring energy from convection into the magnetic field.
A suppressing of convection would then result in the radius inflating to conserve flux.
Though this has been modeled to be possible for stars with radiative cores, modelling of activity-causing inflation for fully convective stars has found that too high a level of activity would be required for observed levels of inflation \citep{Feiden2013b, morales10}.

Stellar activity can also increase uncertainty in our observed results, complicating efforts to define and understand inflation.
The effect of starspots on the measurement of the companion radius by the transit method has been observed on multiple occasions \citep{czesla2009,carter2011}.
When spots move across the visible stellar disc as their star rotates they will create periodic variations in its light which, provided the lifetime of the spot is not short, can be detected.
For longer lifetime active regions there is a possibility to create systematic errors in radius measurement of a size dependent on the strength and number of the active regions.
Their impact is dependent on whether they are occulted by the companion star or not \citep{czesla2009,pont2013prevalence,oshagh2013}. 
Observations of planets eclipsing dark spots as they transit the star have been observed, shown very clearly in the case of HAT-P-11 \citep{southworth2011}, with small peaks during the transit dip being clearly visible in the light curve.
These peaks can cause underestimation or overestimation of the transit depth depending on how they are treated. 
There is also dependence on whether the average surface brightness of the occulted band is less or more than average surface brightness in the non-occulted portions of the star. 
Dark spots not occulted by the companion have a different effect.
The presence of cooler spots on the stellar disc results in the star itself seeming cooler.
This will result in a greater fraction of flux being blocked by a companion and an overestimate of the derived radius.
Starspots can also effect the predictions of stellar models, blocking flux and causing inflation in the pre-main sequence and zero age main sequence that can lead to incorrectly determined ages \citep{spruit,Somers15,Somers20}.

Additionally, in observing these levels of stellar activity there could be an observational bias.
The majority of well-defined low-mass star systems come from short-period binaries. 
Such systems are thought to be tidally locked in synchronised, circular orbits \citep{zahn1977}.
Tidal interactions could increase the speed of the internal stellar dynamo and lead to higher magnetic activity, inhibited convection and thus inflation (e.g. \citealt{ribas2006}). 
However the observation of radius inflation in the case of isolated M-dwarfs (e.g. \citealt{berger2006,boyajian2012,spada2013}) and rapidly rotating low mass stars in binaries without inflation \citep{blake2008}, does suggest a more complicated picture.

Another proposed contributing factor towards the radius inflation problem is metallicity. 
As changes in metallicity results in changes in stellar opacity, it is expected to have a small but noticeable effect on low-mass stellar radii. 
As the outer layers of a star see a decrease in opacity with a lower metallicity, there is a likewise decrease in radiation pressure and therefore in the size of the star. 
This direct effect on a star's structure is accounted for in stellar models, but some studies have suggested a clear trend between inflation and metallicity \citep{berger2006,von2019eblm}.
This would imply that the structural models are not accounting for metallicity correctly, perhaps indicating some missing physics or opacity that causes an underprediction of radius for a fixed mass.
The extent of this effect of metallicity on inflation is debated, with other studies finding no such trend \citep{demory2009}.

To explore the radius inflation problem and address a lack of data for M-dwarfs the Eclipsing Binaries with Low Mass stellar companions (EBLM) project \citep{triaud} was launched.
The EBLM project makes use of the Wide Angle Search for Planets (WASP, \citealt{pollacco2006wasp}) a survey that has found over 150 transiting exoplanets.
WASP also detected a large number of ``false positive'' objects that were detected as ``exoplanet-like'' but whose signals were created in a different way \citep{Schanche2019}.
One of the most common false positives were eclipsing binary stars, which create a similar transit signal as one star orbits the other.
This was especially the case for low-mass stars in eclipsing binaries as their radii, and therefore transit depths, are very similar to those of hot Jupiters.
The EBLM project seeks to use this large source of identified eclipsing binaries to address a shortfall of accurate mass, radius and effective temperature measurements for low mass stars, further exploring apparent problems at the low-mass end of the HR diagram.
The EBLM series has explored eclipsing binaries at different stellar limits \citep{triaud,von2017}, the impact of different models for primary stars \citep{Duck}, potential radius inflation \citep{von2019eblm,gill2019eblm} and in EBLM IV \citep{triaud2017} derived masses from the spectroscopic orbits of over 100 M-dwarfs.

The \textit{CHEOPS} mission \citep{Benz} is the first small (S-class) European Space Agency mission.
Launched on the 18th of December 2019, its primary function is to perform ultrahigh-precision photometry of bright stars known to host exoplanet systems.
The \textit{CHEOPS} guaranteed-time observing (GTO) programme includes ``Ancillary Science'' programmes,  where the targets are not exoplanets but are important to the field of exoplanets.
This includes our programme, ``ID-037 Eclipsing binaries with very low mass stars''.
It seeks to use the capabilities of \textit{CHEOPS} to explore the radius inflation problem.
Additionally, we use data from the \textit{TESS} mission \citep{Ricker} to ensure consistency between different instruments, with different studies reporting inconsistent results for the same object being a previous problem in observing EBLMs \citep[e.g.][]{Yilen,swayne2020tess,martinsub}.

This paper presents the final results for 23 targets from our CHEOPS observing programme.
We focused on targets with masses below the fully-convective boundary (0.35 ${\rm M_\odot}$), as this region is sparsely populated in combined mass, radius and effective temperature measurements.
The first results of the programme were presented in \cite{swayne21}.
Results for 5 systems negligibly affected by star-spot activity were presented in \cite{Sebastian2022}.
These studies demonstrated the capability of \textit{CHEOPS} to provide precise radius and effective temperature measurements for M-dwarfs.
We reanalyse these targets due to the use of new techniques in analysis since \citet{swayne21} and to apply our new methods in starspot simulation to the targets in \citet{Sebastian2022}.
Our targets have had both their primary transit and secondary eclipses observed when possible in both \textit{CHEOPS} and \textit{TESS}.
A few targets only had their secondary eclipse observed with one satellite, but had both their transit and eclipses observed by the other.
In this case we set orbital parameters at those observed by the other satellite and only fit the eclipse depth and orbital shape parameters.
Our observations, data reduction and methods to characterise the host star are outlined in Section~\ref{sec:obs}.
Our analysis of the \textit{CHEOPS} and \textit{TESS} light curves and derivations of the absolute stellar parameters are detailed in Section~\ref{sec:ANA}.
Our approach to account for uncertainties deriving from starspot-induced flux variation is displayed in Section~\ref{sec:spt}
We present our results in Section~\ref{sec:res}, discuss the search for radius inflation trends in Section~\ref{sec:dis}, and give our conclusions in Section~\ref{sec:conc}, commenting on areas of future interest.

%% file: ana.tex
Our \textit{CHEOPS} data analysis follows the methods used in \cite{swayne21} and the analysis of \textit{TESS} light curves follows the methods used in \cite{Sebastian2022}:
\begin{enumerate}
    \item From transit photometry obtain transit and secondary eclipse depths (allowing us to calculate radius ratios and flux ratios), surface gravity and the stellar density when combined with mass from the \textit{TESS} input catalogue. The \textit{TESS} photometry also allows us to derive orbital periods from all but our longer period systems.
    \item Combine density with our primary effective temperatures and metallicities to calculate the primary stellar mass using the equation from \citet{enoch10}.
    \item Using density and mass, derive primary stellar radius.
    \item Through the primary stellar mass and derived mass functions we calculate secondary stellar mass. Through primary stellar radius and derived radius ratio we calculate secondary stellar radius.
    \item We iterate once again through steps (ii) - (iv) to ensure mass ratios are consistent with the stellar density.
    \item Using the primary stellar parameters we derive a theoretical surface brightness. Through using the observed flux ratio and transit depth we can thus derive the surface brightness of the secondary star. Combining with stellar parameters, we can derive the effective temperature of the secondary star.
    \item Finally, through generating theoretical radii and effective temperatures for the given masses of our stars, we can calculate how our observed results deviate.
\end{enumerate}
This approach uses the posterior probability distributions (PPDs) of each parameter to accurately calculate the uncertainties of our correlated errors as we derive our results.
Light curves in both methods were modelled using the \texttt{qpower2} algorithm to compute light curves with the power-2 limb darkening law \citep{maxted2019q}.
This is used in a binary star model of both primary and secondary eclipses present in \texttt{PYCHEOPS}, the data analysis \textsc{PYTHON} package purpose-built for the \textit{CHEOPS} mission \citep{maxted21}.
The limb-darkening effect is applied to the primary eclipses whereas the secondary eclipses are modeled assuming a uniform stellar disc for the secondary star as limb-darkening has a negligible effect on the light curve.
For the purpose of describing the models, parameters $R_1$ and $R_2$ are the radii of the primary and secondary stars respectively.
The parameters used in the binary star model are: the time of mid-primary eclipse $T_0$; the primary transit depth parameter $D = {R_2}^2/{R_1}^2$, the impact parameter $b = a \cos{i}/R_1$, where $i$ is the orbital inclination and $a$ is the semimajor axis; the transit width parameter $W = \sqrt{(1 + R_2/R_1)^2 - b^2} R_1/(\pi a)$; the eccentricity and argument of periastron dependent parameters $f_s = \sqrt{e} \sin{(\omega)}$ and $f_c = \sqrt{e} \cos{(\omega)}$; the secondary eclipse depth $L$ and the limb-darkening parameters for the primary star $h_1$ and $h_2$ as defined by \cite{maxted2018}.
$D$ is the transit depth in absence of limb darkening while $W$ is the  transit width in phase units assuming a circular orbit, parameterising the transit depth and width. 

For some of our targets, we only obtained one primary transit and one secondary eclipse event with \textit{CHEOPS}.
Therefore, we had to fix the orbital period to a known value.
For consistency we did this for all our \textit{CHEOPS} analyses.
For every target we fixed $P$ to the value obtained by our analyses of the \textit{TESS} light curves, with the exception of EBLM~J1559-05 which had no \textit{TESS} light curve and EBLM~1928-38 where \textit{TESS} does not observe primary or secondary occultations.
The orbital period of EBLM~J1559-05 was set at a value obtained from \citet{triaud2017} and EBLM~J1928-38 at a value obtained from \citet{martin2019}.
For those of our targets in zero eccentricity systems we set $f_c$ and $f_s$ to be at a constant value of zero assuming a circular orbit.
For our eccentric systems we set priors on $f_c$ and $f_s$ based on the obtained or derived values of eccentricity and arguments of periastron.
Additionally, priors in $h_1$ and $h_2$ were included for the EBLMs J0719+25, J1741+31 and J1934-42.
The values used for the priors were derived using interpolation in the  data tables presented in \citet{maxted2018} based on the limb-darkening profiles from the STAGGER-grid \citep{magic2015stagger}.
The interpolation is performed based on the effective temperature, surface gravity and metallicity from Table \ref{Primparams}.
An offset (0.01 for $h_1$, $-$0.045 for $h_2$) was then applied based on the offset between empirical and tabulated values of these limb darkening parameters observed in the \textit{Kepler} bandpass by \citet{maxted2018}.
For EBLM~J0719+25 we used a prior as $h_2$ was trending to unphysically low values if left without a prior.
For EBLM~J1741+31 and EBLM~J1934-42 we used priors due to the same reason as in \cite{swayne21}, as the partial primary eclipses did not put enough constraint on the limb darkening parameters.

\begin{figure*}
	\includegraphics[width=\linewidth]{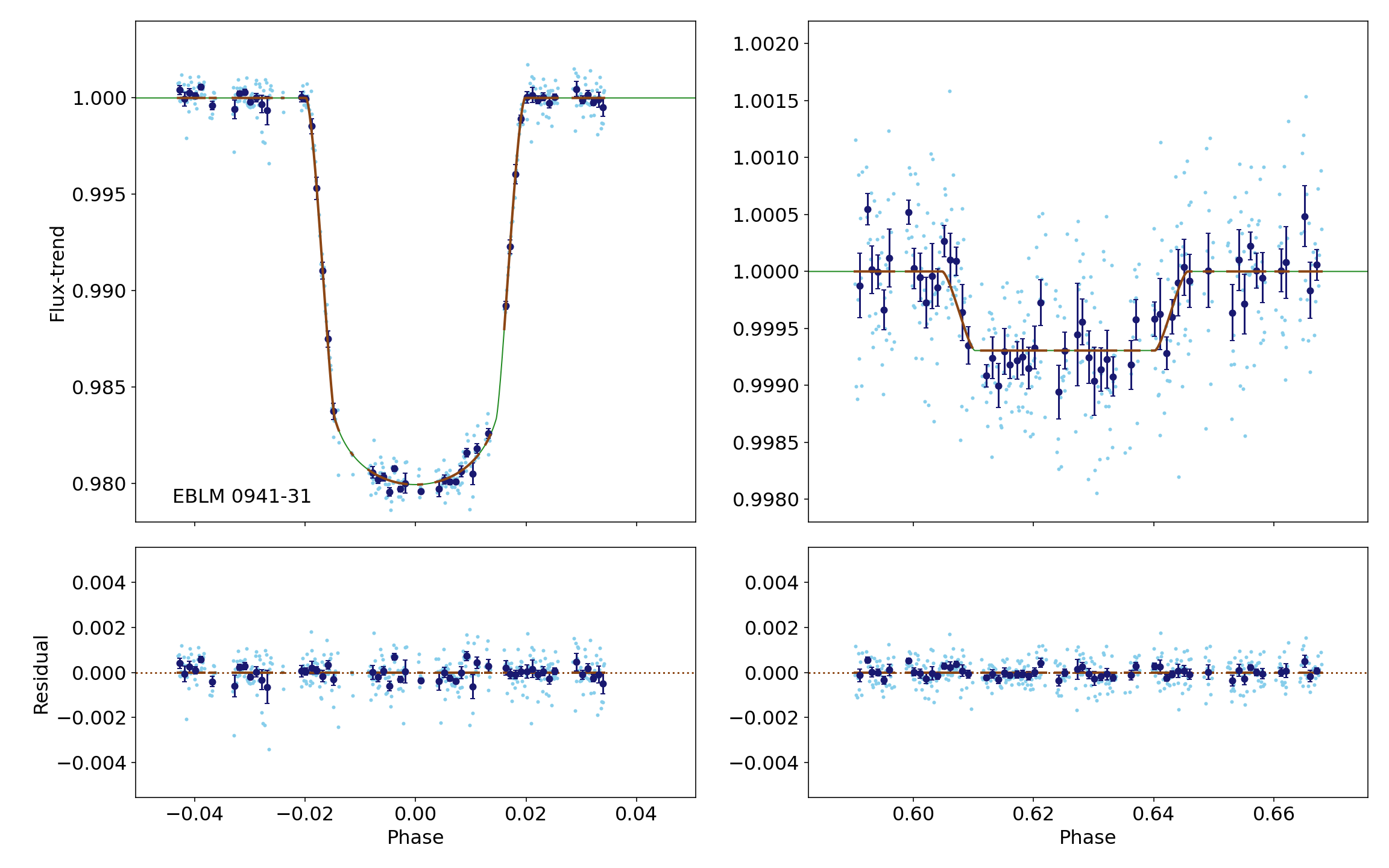}
    \caption{Fitted \textit{CHEOPS} light curve of  EBLM~J0941-31 in phase intervals around the primary and secondary eclipse events. The observed data points are shown in cyan. The transit and eclipse models are shown in green. Binned data points with error bars are shown in blue and the fit between binned data points in brown. The residual of the fit is displayed below the fitted curves.}
    \label{fig:fCHEOPS_lcs4}
\end{figure*}

\subsection{Analysis of \textit{CHEOPS} light curves}

Our \textit{CHEOPS} light curves from each visit were first analysed separately to derive initial model parameters and choose decorrelation parameters.
A log of each visit can be found in Table \ref{FullObsLog}.
Firstly, we determined initial orbital parameters with a least-squares fit.
As fully detailed in \citet{swayne21}, instrumental effects such as satellite roll angle or contamination can be modelled for using linear decorrelation parameters or with roll angle $\rm \phi, sin(\phi), cos(\phi), sin (2\phi)$, etc.
These can be selected iteratively over a number of least-squares fits by calculating their Bayes factors and discarding those parameters with the largest factors until $B_p>1$ for all remaining parameters as discussed in Section 3.4 of \citet{maxted21}.
We then sampled the PPD of the model and decorrelation parameters simultaneously using the Markov chain Monte Carlo (MCMC) code \texttt{EMCEE} \citep{Foreman}.
For primary eclipse fits we used Gaussian priors for $\rm f_c$, and $\rm f_s$ and set orbital periods to values obtained from \textit{TESS} or from radial velocity fits.
For secondary eclipse fits we also used priors on $D$, $W$ and $b$ based on the values derived from primary transit fit of the target.

After this initial step, the single visits for each target were combined in a ``MultiVisit'' analysis.
These analyses of all \textit{CHEOPS} visits for a target used the same priors as the individual fits and used a combined average of their derived results as input parameters.
Using the \texttt{MultiVisit} function of \texttt{PYCHEOPS}, \texttt{EMCEE} could be used to sample the joint PPD of each target.
Each individual visit decorrelation parameter selected in the initial fits were included in the sampling with the exception of roll angle which was calculated implicitly using the method described in \cite{maxted21}, with the number of harmonic terms kept to the default value nroll=3.
An example of the fitted light curve is shown in Figure \ref{fig:fCHEOPS_lcs4}.

\subsection{Analysis of \textit{TESS} light curves}

As in \citet{Sebastian2022} we removed trends in the \textit{TESS} light curve with a different method to that used in \citet{swayne21}.
Only segments of the \textit{TESS} light curve within one eclipse duration of the mid-eclipse were used in the analysis.
These segments were divided by a linear polynomial fitted to the data either side of the masked-out eclipse event.
To model the light curve, we again used a least-squares fit to obtain initial model parameters before sampling the PPD of our \texttt{PYCHEOPS} EBLM model using \texttt{EMCEE}. 
Normal priors were placed on the orbital parameters $\rm f_c, f_s$ using the same values as used in the \textit{CHEOPS} fitting as well as on the white noise, using the residual rms of the least-squares fit.

\subsection{Derivation of primary stellar mass and radius}

To obtain the primary stellar mass, we use the empirical relation  M(${\rm T}_{\rm eff}$, $\rho$, [Fe/H]) from \citet{enoch10}.
Values of radius are then calculated from mean stellar density of the primary star.
With effective temperature and metallicity derived by the TS3 (Tech and Support 3) - Target Characterisation working group of \textit{CHEOPS}, the only further quantity we needed to use the equations was the stellar density $\rho$.
We first used estimates of primary mass and radius from the \textit{TESS} input catalogue v8 \citep{stassun2019revised} as initial parameters; along with RV semi amplitude $K_1$ and orbital parameters $e$ and $\omega$ from RV measurements; and radius ratio $k$, semi-major axis divided by stellar radius $a/R_1$ and the sine of the inclination $\sin{(i)}$ from our transit observations.
With the {\tt MASSRADIUS} function of \texttt{PYCHEOPS}, we used the estimates of stellar mass and radius to get an estimate sample of mean stellar density for our sample of fitted transit parameters using:
\begin{equation}
\langle \rho_{\star}\rangle = 3\pi a^3/(GP^2(1+q)R_1^3)
\end{equation}
where $q$ is the mass ratio $M_2/M_1$ and $G$ is the gravitational constant.
{\tt MASSRADIUS} propagates errors using a Monte Carlo approach with a sample of 100,000 points per parameter, with the mean and standard error of input values used to generate normal distributions or alternatively using samples taken directly from MC distributions.
In our case the density uses input values of $M_1$, $P$ and $K_1$ and the samples from the MCMC fit of $R_1/a$ and ${\rm sin(i)}$.
With the calculated density we derived a mass sample for the primary star with the equation for mass from \citet{enoch10}, using normal distribution samples of effective temperature and metallicity based on the values and uncertainties derived by the TS3 working group.
After adding a normal distributed scatter of 0.023 in the logarithm of the mass to account for the scatter in this relation reported by \citet{enoch10}, we used the mass and density samples to derive a radius sample.
These would be the primary stellar samples used in the final calculations of secondary stellar mass and radius.

\subsection{Deriving secondary stellar mass, radius and effective temperature}

We calculated secondary stellar mass and radius using \texttt{MASSRADIUS}.
The function uses the PPD of the light curve fits to derive the mass and radius of the companion star from the analysis of the light curve, given an estimate for the primary star's mass and $K_1$, as explained in \citet{maxted21}.
With the primary stellar parameters and those orbital parameters not derived from our light curves presented in Section \ref{sec:obs} in Table \ref{Primparams}, we could derive the secondary star's mass and radius as well as the surface gravities of both bodies


We derived the effective temperature T$\rm _{eff,2}$ of the M-dwarf companion using the surface brightness ratio $\rm L/D$, derived from our fits to the primary and secondary eclipse.
Using the same approach as detailed in \citet{swayne21}, we made use of PHOENIX model atmospheres with no alpha-element enhancement \citep{Husser} for different spectral parameters T$\rm _{eff}$, $\rm log\,g$, and [Fe/H] to create a grid of theoretical integrated surface brightness in the \textit{CHEOPS} and \textit{TESS} passbands.
With the known primary stellar parameters we derived a sample of surface brightness values for the primary star.
Using the observed surface brightness ratio, the metallicity (assuming similar metallicity for both stars) and the surface gravity of the secondary star, we used bisection to obtain a PPD of secondary stellar effective temperature for each target.
To estimate the systematic error in these values of T$\rm _{eff,2}$, we compared the integrated surface brightness computed with the models of a \citeauthor{Husser} to a variety of stellar atmosphere models obtained from the Spanish Virtual Observatory Theoretical Model Service\footnote{\url{http://svo2.cab.inta-csic.es/theory/main/}}. 
This comparison was done  for models with T$\rm _{eff}=3000$\,K, $\rm log\,g = 5.0$, and ${\rm [Fe/H] } = 0.0$. 
The integrated surface brightness in the CHEOPS passband varies from 0.92\,per~cent to 1.18\,per~cent relative to the surface brightness integrated over all wavelengths. 
In the TESS passband, the range is from 1.72\,per~cent to 2.12\,per~cent. 
This corresponds to a systematic error of about 50\,K in our estimates of T$\rm _{eff,2}$. 

%% file: spt.tex
For six of our targets, the \textit{TESS} light curve shows clear pseudo-periodic variations in flux on a timescale of a few days.
This same effect can be seen as gradients with time in the \textit{CHEOPS} data.
This variation is similar to that reported elsewhere for eclipsing binaries \citep{Sethisub} and is due to starspots.
In the simplest scenario this involves the dipping of the level of flux as a starspot travels from one side of the stellar disc to the other, resulting in a curved dip of light due to the change in the area projected on the disc by the spot and the effect of stellar limb darkening.
The presence of multiple evenly spaced spots on a rotating star could thus create what appears to be a periodic sinusoidal signal.
In reality, a combination of starspots of differing sizes, positions and even varying period makes the signal more complicated than an actual sinusoid.
As the effect of starspots on transit observations can result in both overpredicted or underpredicted stellar radii and we are exploring the radius inflation problem, we decided to build a series of functions in \textsc{PYTHON} to quantify the effect of stellar activity for each of our objects as an uncertainty, to be added to our final radius results.

\subsection{Fitting the starspot signal}

\begin{figure*}
    \includegraphics[width=\linewidth,height=\linewidth]{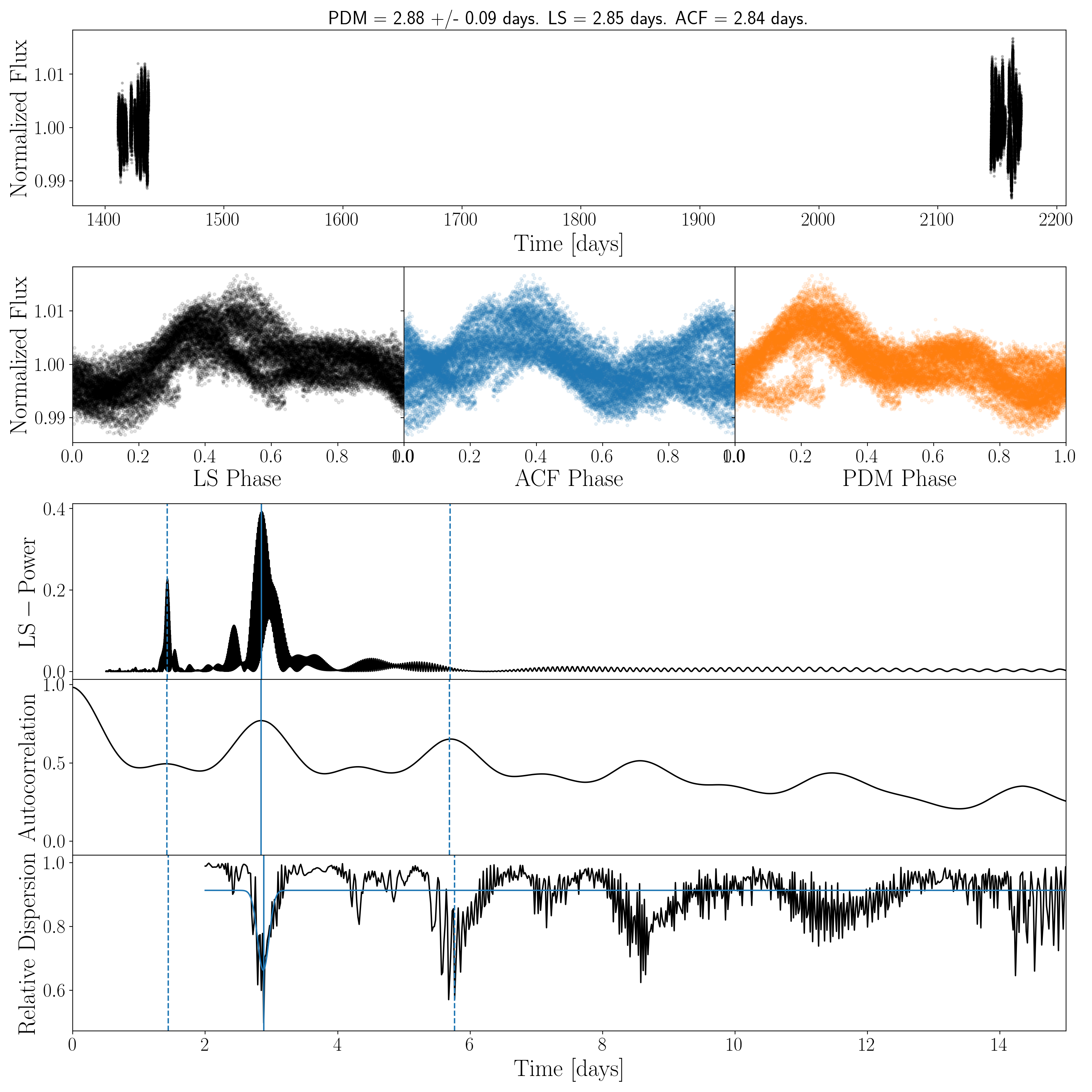} 
    \caption{A set of output plots generated by the module {\tt STARSPOT} when analysing the masked flux signal of EBLM J0239-20. 
    The top plot displays the inputted flux signal. 
    The second shows the flux signal phase folded by the fitted variation period for each method (Lomb-Scargle, autocorrelation functions and phase dispersion minimisation). 
    The third, fourth and fifth plots are the plotted results of each method showing the period versus the signal detection likelihood. 
    In this example Lomb-Scargle periodograms finds a variation signal with a period of 2.85 days, the autocorrelation function finds a variation signal with a period of 2.84 days and the phase dispersion minimisation fits a period of 2.88 days.}
    \label{fig:0239spot}
\end{figure*}

We first sought to measure the observed stellar activity for each of our targets.
With \textit{TESS} light curves we had sources of long continuous light curves for nearly all targets.
In order to obtain the stellar rotation periods from the light curves we used the {\tt STARSPOT} package\footnote{https://github.com/RuthAngus/starspot}.
{\tt STARSPOT} is a \textsc{PYTHON} module designed to obtain the stellar rotation period using autocorrelation functions, Lomb-Scargle periodograms and phase dispersion minimisations.
We masked out the transit and secondary eclipses of the light curve so it was purely fitting the activity signal.
An example of the flux signal and its analysis by {\tt STARSPOT} is shown in Figure \ref{fig:0239spot} for the EBLM~J0239-20.
In general we found that only the Lomb-Scargle periodogram obtained a definitive and clear period for the variation signal.
Therefore, it was used as the method to obtain our variation periods.

In order to characterise the stellar signals we decided to fit them with a sinusoidal function with one harmonic:
\begin{equation} \label{activityeq}
F(t) = C + a_1 \sin{(2\pi t/P_{\rm rot} + \phi_1)} + a_2 \sin{(4\pi t/P_{\rm rot} + \phi_2)}
\end{equation}
where $a_1$ and $a_2$ are the amplitudes of the stellar activity signal, $\phi_1$ and $\phi_2$ are phase constants, $C$ is a constant, $t$ is time and $P_{\rm rot}$ is the period of the stellar  activity signal.
With the period of the stellar activity signal fixed at the value obtained from the {\tt STARSPOT} analysis, we fit the function using the curve\_fit function of {\tt SCIPY}.
We split the light curve into slices 5000 data points wide, covering around a sixth of a typical \textit{TESS} sector observation.
We then created a new 5000 point wide slice for every 2500 points giving us around 11-13 overlapping slices for each \textit{TESS} sector observation.
We would then find the amplitude of the variation for each slice and obtain the mean amplitude, using the standard deviation as the range of stellar activity variation shown by the target star.

\subsection{Simulating spot patterns}

To quantify the effect of starspots on fitted orbital parameters, we simulated starspot perturbed light curves for each  EBLM system, and then performed the same fit as we would upon our observed light curves.
Any changes in the observed orbital parameters would thus be caused by the introduced stellar activity signal.
To do this we used the \textsc{PYTHON} module {\tt ELLC} \citep{maxted2016ellc} as it has the ability to include starspots in its light curve model.
{\tt ELLC} uses integrals from \citet{EkerA,Ekerb}, expressing how circular spots affect the light curve of a spherical star with quadratic limb darkening to calculate spot-induced flux variation for its model light curve.
The effect of spot-crossings during eclipse is also accounted for in the model.
How these effects are applied can be found in Section 2.10 of \citet{maxted2016ellc}.
However, as {\tt ELLC} introduces spots via user-selected longitude, latitude, size and brightness factor (the brightness of the spot relative to the local photosphere) there is no direct way to gain an activity signal of the desired amplitude. 
Therefore, we needed to generate spot patterns capable of causing the observed amplitude of a target's activity.
We decided to do so using the Sun as a basis in constructing realistic spot patterns.
The Sun is easily the most observed and documented example of spot activity upon stars and other stars have been reported as following the same spot patterns as the Sun but with varying activity levels \citep[e.g. HAT-P-11,][]{Morris17}.
As our targets covering a range from F9 to K2 stars, we approximate them to have similar spot pattern behaviour to the Sun.

During its most active phases the Sun can have a few hundreds of spots present on its surface at a single time \citep{Clette}.
\citet{giles2017} note that the modulation of solar photometric variability is dominated by the largest individual active regions.
We therefore looked at the number of spot groups instead, making the simplification that the activity caused by a large number of spots in a single group was equal to that caused by a single spot of greater size, and that these large ``spots'' dominate our observed stellar activity signals.
This greatly reduces the time required to compute the synthetic light curves.
We thus looked at the group number statistics provided by the SILSO (Sunspot Index and Long-term Solar Observations) world data centre\footnote{https://www.sidc.be/silso/}.
Using their archives of daily sunspot group numbers \citep{Hoyt1,Hoytb,Vaquero}, we looked for the number of spot groups present at times of maximum solar activity.
The highest numbers are found to be from 10-16 spot groups, so we used this as our distribution of spot group numbers. 
We use the Sun at maximum solar activity as a proxy for the typical variability due to spots seen in our targets, as the variation in total solar irradiance due to rotation near solar maximum is about 0.1\,per~cent up to about 1\,per~cent \citep{2004A&A...414.1139A}, which is comparable to the range of amplitudes due to spots and rotation that we can measure in \textit{TESS} light curves.

To work out an appropriate area for our group-representing spot we turned to the work of \citet{Baumann}.
They find that the group area is well described by a variety of fitted log-normal distributions with $\langle A \rangle$ being the mean area and $\sigma_A$ being the width of the log normal distribution. 
When testing the function we took as values for mean area and distribution width values of 62.2 and 2.45 micro solar hemispheres, respectively, from the ``Total Area'' dataset in Table 1 of \citet{Baumann}.
Thus we generated spot group areas from this log-normal distribution for however many ``spots'' we needed.
We have targets with a greater activity level than the Sun, so we introduced a factor $A_{\rm fac}$ to increase the chosen spot areas depending on the observed amplitude of the spot signal.
This factor is generated before the spot pattern itself based on the inputted activity amplitude and uses the bisection method to roughly narrow-in on an appropriate $A_{\rm fac}$ with the decision on which bisected segment to take depending on the activity amplitude generated.

To generate astrophysically sensible spot positions we used the work of \citet{Hathaway}.
\citet{Hathaway} defines two equations describing spot position firstly the active spot latitude:
\begin{equation} \label{actlat}
\overline{\lambda}(t) = 28^{\circ} \exp{[-(t-t_0)/90]}
\end{equation}
where $\overline{\lambda}$ is the active latitude, $t_0$ is the starting time of the solar cycle and $t$ is the current time in the cycle with both times in months.
The second equation described the latitudinal width of the sunspot zones, finding a relation for the RMS of the width of their sunspot zones as:
\begin{equation}\label{sptband}
\sigma_{\lambda}(A) = 1.5^{\circ} + 3.8^{\circ}(1-\exp{[-A/400]})
\end{equation}
where $\sigma_{\lambda}(A)$ is the RMS width of the sunspot zone and $A$ is the total sunspot area in micro hemispheres.
As we are taking our standard quantities in area to be at times of maximum activity we also applied this to the spot latitude.
We used Figure 43 of \citet{Hathaway} to approximate the time of maximum activity in the solar cycle to be 50 months.
Thus in deciding the active latitude in Equation (\ref{actlat}) we set $t$ to be 50 months.
We rejected any spot patterns that included any overlapping spots from our simulations.

Through $A_{\rm fac}$ our spots would be generated around the right range but with our random distributions we made sure that our sample would avoid bias.
Due to this we still generate patterns which resulted in stellar variability completely different to our observed signal.
These are discounted by fitting the generated light curve for Equation (\ref{activityeq}) and only recording the pattern of those with amplitudes within the standard deviation range of our observed amplitudes.
We run the routine until the required number of acceptable patterns are generated.
We found that this gave an almost uniform distribution of amplitude $a_1$ and an amplitude $B$ tending slightly towards lower values.
Therefore we concluded that we were sufficiently unbiased for our simplified method.

With our routine we generate 500 unmasked light curves, which gives a good balance between computation time and sample size.
We then fit these light curves with {\tt PYCHEOPS} using the least squares fit that we use to initially fit \textit{TESS} light curves.
Using how they differ from what is input, we can quantify the impact the stellar activity and removal of it has had on our retrieval of the system's characteristics.
We applied our routine to all our targets with \textit{TESS} light curves.
For EBLM~J1559-05, the only target without a \textit{TESS} light curve, an analysis of it's \textit{WASP} light curve were performed using the method in \citet{Maxted2011} to obtain the amplitude and period of any present variation.
An upper limit of 2mmag was found for the system.
However, as the period of the rotation signal is close to the orbital period it is not clear if the signal truly is for rotation.
As no consistent rotation period was found, we fit for a variation of period 10.5 days but will not apply the derived corrections to our final results.
For targets J0239-20, J1928-38 and J2040-41; we fixed orbital period $P$ and orbital parameters $f_c$ and $f_s$ in the least squares fits.
This was due to the least square fitting having difficulty detecting the very small inputted eclipse depths, leading to very large uncertainties in ${\rm T}_{\rm eff,2}$
The effect on radius predicted by our starspot-induced variation is shown in Table \ref{spotresults}.

\begin{table*}
    \centering
      \caption{The details and results of our starspot simulations. Var. Period is the period of observed variation in normalised flux, Var. Amplitude is the observed amplitude of the stellar variation in normalised flux for each of our targets. The originally input values of radius and effective temperature are listed alongside the resulting radius and effective temperature and the resultant change in radius and effective temperature induced by the spot patterns.}
         \label{spotresults}
    \resizebox{\textwidth}{!}{$ 
         \begin{array}{lrrrrrrrr}
            \hline
            \hline
            \noalign{\smallskip}
             \text{Target}  & \multicolumn{1}{c}{\text{Var. Period (days)}} & \multicolumn{1}{c}{\text{Var. Amplitude}}  & \multicolumn{1}{c}{\rm R_{\rm 2,input}\, ({\rm R_\odot})} & \multicolumn{1}{c}{{\rm R}_{\rm 2,output}\, ({\rm R_\odot})} &
             \multicolumn{1}{c}{\rm \Delta R \, (\%) } & \multicolumn{1}{c}{{\rm T}_{\rm eff,2,input}\, ({\rm K})} & \multicolumn{1}{c}{{\rm T}_{\rm eff,2,output}\, ({\rm K})} & \multicolumn{1}{c}{\Delta {\rm T}_{\rm eff,2}\, ({\rm K})} \\
            \noalign{\smallskip}
            \hline            \hline
            \text{J0057-19} & 4.94 & 0.0057 \pm 0.0022 & 0.1668 & 0.1651 \pm 0.0053 & 1.04 & 2958 & 2990 \pm 57 & 32 \\
            \text{J0113+31} & 18.11 & 0.0014 \pm 0.0004 & 0.2152 & 0.2163 \pm 0.0041 & 0.51 & 3258 & 3262 \pm 24 & 4 \\
            \text{J0123+38} & 5.74 & 0.0036 \pm 0.0008 & 0.3424 & 0.3410 \pm 0.0100 & 0.42 & 3404 & 3414 \pm 87 & 10  \\
            \text{J0239-20} & 2.85 & 0.0049 \pm 0.0017 & 0.2022 & 0.2048 \pm 0.0055 & 1.27 & 3027 & 3054 \pm 266 & 27 \\
            \text{J0540-17} & 6.50 & 0.0005 \pm 0.0002 & 0.1917 & 0.1928 \pm 0.0047 & 0.59 & 3220 & 3236 \pm 26 & 16\\
            \text{J0546-18} & 3.32 & 0.0021 \pm 0.0004 & 0.2194 & 0.2209 \pm 0.0094 & 0.70 &3412 & 3429 \pm 40 & 17 \\
            \text{J0719+25} & 5.24 & 0.0018 \pm 0.0009 & 0.1847 & 0.1859 \pm 0.0055 & 0.64 & 3212 & 3200 \pm 73 & -12\\
            \text{J0941-31} & 5.28 & 0.0013 \pm 0.0006 & 0.2286 & 0.2286 \pm 0.0060 & 0.02 & 3448 & 3434 \pm 39 & -14 \\
            \text{J0955-39} & 27.79 & \text{Not fittable} & \text{---} & \text{---} & \text{---} &\text{---}&\text{---}&\text{---} \\
            \text{J1013+01} & 3.3 & 0.029 \pm 0.009 & 0.2100 & 0.2112 \pm 0.0041 & 0.56 &3043 &3036 \pm 33 & -7 \\
            \text{J1305-31} & 4.89 & 0.0010 \pm 0.0003 & 0.2986 & 0.2993 \pm 0.0068 & 0.23 & 3135 & 3131 \pm 20 & -4 \\
            \text{J1522+42} & 7.58 & 0.0008 \pm 0.0006 & 0.1888 & 0.1898 \pm 0.0042 & 0.53 & 3073 & 3070 \pm 21 & -3\\
            \text{J1559-05} & \text{--} & 0.001 & 0.1977 & 0.1984 \pm 0.0043 & 0.36 & 3139 & 3161 \pm 33 & 22 \\
            \text{J1741+31} & 7.64 & \text{Not fittable} & \text{---} & \text{---} & \text{---} &\text{---}&\text{---}&\text{---}\\
            \text{J1928-38} & 13.25 & 0.0009 \pm 0.0006 & 0.2672 & 0.2670 \pm 0.0054 & 0.06 & 3153 & 3155 \pm 21 & 2 \\
            \text{J1934-42} & 4.21 & 0.0032 \pm 0.0011 & 0.2244 & 0.2256 \pm 0.0063 & 0.54 & 3014 & 3317 \pm 770 & 303 \\
            \text{J2040-41} & 14.20 & 0.0010 \pm 0.0008 & 0.1766 & 0.1755 \pm 0.0061 & 0.64 & 2910 & 2924 \pm 19 & 14 \\
            \text{J2046-40} & 14.66 & 0.0028 \pm 0.0030 & 0.2196 & 0.2207 \pm 0.0046 & 0.51 & 3163 & 3163 \pm 38 & 0\\
            \text{J2046+06} & 10.94 & 0.0004 \pm 0.0002 & 0.2034 & 0.2037 \pm 0.0041 & 0.17 & 3124 & 3126 \pm 24 & 2 \\
            \text{J2134+19} & 18.05 & 0.0012 \pm 0.0011 & 0.3691 & 0.3666 \pm 0.0086 & 0.69 & 3496 & 3488 \pm 30 & -8\\
            \text{J2315+23} & 5.21 & 0.0012 \pm 0.0002 & 0.2465 & 0.2481 \pm 0.0067 & 0.65 & 3298 & 3297 \pm 24 & 1\\
            \text{J2343+29} & 9.57 & \text{Not fittable} & \text{---} & \text{---} & \text{---} &\text{---}&\text{---}&\text{---}\\
            \text{J2359+44} & 4.37 & 0.0009 \pm 0.0002 & 0.2942 & 0.2948 \pm 0.0067 & 0.22 & 3462 & 3496 \pm 93 & 34 \\
            \hline
         \end{array}
    $}
    \medskip
\end{table*}

For systems with high flux variation there is a small change to the derived radius, introducing variations in secondary radius at the sub-1\% level.
When there is less variation there is generally less of a change in radius.
There are also small variations in effective temperature that seem to roughly increase with increased variation, again mostly at the sub-1\% level.
This is the expected result and showed that our method can provide a reasonable estimate for the variation in radius and effective temperature caused by the effect of starspots.
One exception is EBLM~J1013+01 which has by far the greatest variation amplitude but whose radius is not mischaracterised by a larger amount than the rest of our sample.
One future area of interest would be to characterise EBLMs with similar flux variation to observe if this error ``cut-off'' is repeated.
Targets EBLM~J0955-39, EBLM~J1741+31 and EBLM~J2343+29 found no rotation signal and will also receive no starspot-derived corrections to radius and effective temperature.
There are two targets with large uncertainties in effective temperature in the fit.
EBLM~J1934-42 has partial eclipses and so the effective temperature we derive is not reliable.
For EBLM~J0239-20, we propose that the large uncertainty in effective temperature is due to the combination of large variation amplitude and period, leading to further difficulty detecting the very small eclipse depth.
We use our starspot results to account for the uncertainty caused by the variation in stellar flux in both \textit{CHEOPS} and \textit{TESS} light curves.
For targets with a detectable rotation signal the uncertainties from our MCMC fits are combined in quadrature with the uncertainties predicted by our starspot simulations.
This was done rather than a correction due to the uncertainties in derived values being larger than any potential applied correction.
This would risk the potential undercorrection or overcorrection observed in previous literature and could be sensitive to varying flux variation amplitudes over large periods of time between \textit{TESS} visits or sectors.
Applying the correction as an additional uncertainty represents the extra uncertainty in derived properties without potentially applying bias to our results.

%% file: res.tex
\section{The results of our programme}
\label{sec:res}

\subsection{Photometry results}

We derived orbital properties using \textit{CHEOPS} and \textit{TESS} results, applying starspot corrections for radius and effective temperature values and combining the two results.
For \textit{TESS} light curves the same priors as applied to the \textit{CHEOPS} light curves were applied with the exception that EBLM~J0719+25 had no limb darkening priors applied as they were not needed.
The absolute parameters of our targets are shown below in Table \ref{Results}.
The primary stellar mass and radii were derived using the equations in \protect\citet{enoch10} as described in Section \ref{sec:ANA}.
We use the secondary stellar masses from the \textit{CHEOPS} fit as our photometry-derived results will have little impact on them.
Where we have fit light curves from both \textit{CHEOPS} and \textit{TESS}, we combine the secondary stellar radii, secondary stellar effective temperature and both primary and secondary surface gravities.
The fit parameters and other derived properties will be available online as supplementary material.

For the targets EBLM~J1522+42 and EBLM~J2046-40, we only obtained visits of the secondary eclipse with \textit{CHEOPS}.
Therefore, we set the orbital parameters $D$, $W$ and $b$ to the values obtained from \textit{TESS} light curves, set $T_0$ as well as $P$ constant and only derive the secondary eclipse depth $L$, $f_c$ and $f_s$ from these visits.
For EBLM~J2134+19 where the two transit visits miss the ingress and egress of the occultation respectively, a Gaussian prior on the orbital period was applied based on data from \textit{WASP} light curves.
For the targets EBLM~J2134+19 and EBLM~J2343+29 the \textit{TESS} light curves miss the primary eclipse.
For these objects, similarly to our \textit{CHEOPS} targets with only the secondary eclipse, we only fit for $L$, $f_c$ and $f_s$ fixing all other parameters at \textit{CHEOPS} values, though for EBLM~J2134+19 we also fit for $W$ as the eclipse was wider than fitted in \textit{CHEOPS}.
\textit{TESS} light curves were not analysed for EBLM~J1559-05 which has not been observed at the time of writing and for EBLM~J1928-38 whose observations missed both primary and secondary eclipses.
The fitted \textit{CHEOPS} and \textit{TESS} light curves for each target will be available online as supplementary material.

\begin{figure}
    \includegraphics[width=\linewidth]{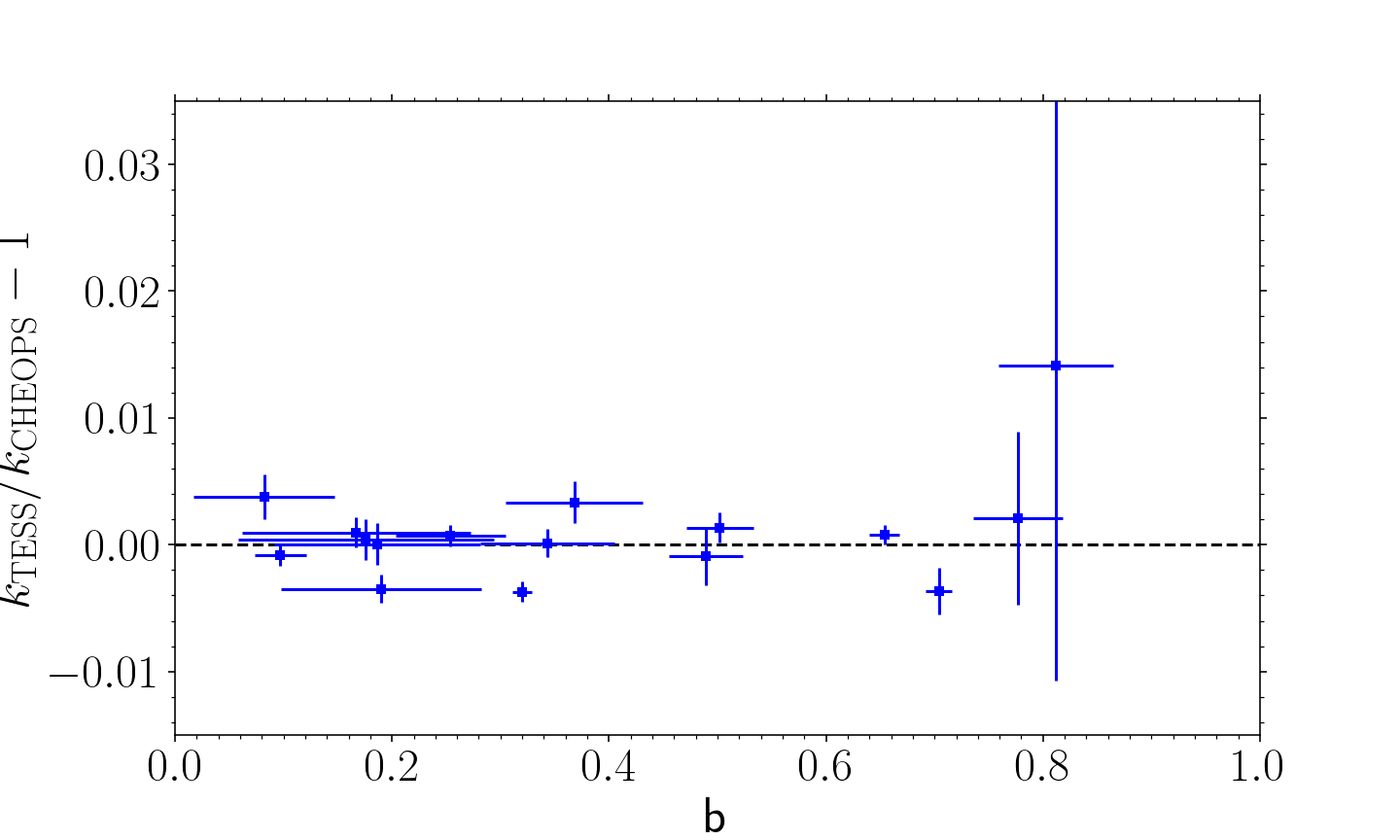} 
    \caption{\textit{CHEOPS} impact parameter versus the difference in observed radius ratio between our \textit{CHEOPS} and \textit{TESS} analyses.}
    \label{fig:CHvTess}
\end{figure}

For the majority of our targets, the \textit{TESS} results agree with the \textit{CHEOPS} results within the bounds of their uncertainty.
We illustrate this by showing the difference in radius ratio between the \textit{CHEOPS} and \textit{TESS} results in Figure \ref{fig:CHvTess}.
The observed differences remains negligible with increased uncertainty for our more grazing systems at higher impact parameters.
There are one or two outliers, that are more than one or two standard deviations away.
However, this is acceptable given 2-sigma confidence levels for a sample of our size (i.e. in a sample of 20, around 1 should fall outside 95\% confidence). 
We expect these differences to be some form of ``analysis noise'', where differences in data reduction such as contamination corrections or background subtraction cause systematic errors.
The difference in fractional primary radius ($R_1/a$) is consistent between instruments.
This consistency between wavelength regimes is a good check of the accuracy of our results and shows that in the case of bad SNR for any \textit{TESS} light curves, that \textit{CHEOPS} light curves can provide data of required precision.

\begin{table*}

      \caption{The absolute stellar parameters derived from our light curve fits for all EBLM targets.}
     \label{Results}
    $ 
         \begin{array}{lrrrrrrr}
            \hline\hline
            \noalign{\smallskip}
            \multicolumn{1}{c}{\text{Target}} & \multicolumn{1}{l}{M_1} & \multicolumn{1}{l}{R_1} & \multicolumn{1}{l}{M_2} & \multicolumn{1}{l}{R_2} & \multicolumn{1}{l}{\log g_1} & \multicolumn{1}{l}{\log g_2} & \multicolumn{1}{l}{{\rm T}_{\rm eff,2}} \\
            & \multicolumn{1}{l}{[{\rm M_\odot}]} & \multicolumn{1}{l}{[{\rm R_\odot}]} & \multicolumn{1}{l}{[{\rm M_\odot}]} & \multicolumn{1}{l}{[{\rm R_\odot}]} & \multicolumn{1}{l}{[{\rm cgs}]} & \multicolumn{1}{l}{[{\rm cgs}]} & \multicolumn{1}{l}{[{\rm K}]}  \\
            \noalign{\smallskip}
            \hline
            \noalign{\smallskip}
            \text{J0057-19} & 1.004 \pm 0.063  & 1.234 \pm 0.037 & 0.1290 \pm 0.0052 & 0.1705 \pm 0.0033 & 4.254 \pm 0.011 & 5.087 \pm 0.012 & 2822 \pm 83\\
            \text{J0113+31} & 1.033 \pm 0.057 & 1.432 \pm 0.027 & 0.1974 \pm 0.0068 & 0.2193 \pm 0.0033 & 4.138 \pm 0.007 & 5.055 \pm 0.009 & 3243 \pm 37\\
            \text{J0123+38} & 1.156 \pm 0.065 & 2.018 \pm 0.055 & 0.338 \pm 0.012 & 0.3531 \pm 0.0060 & 3.885 \pm 0.012 & 4.874 \pm 0.011 & 3479 \pm 60\\
            \text{J0239-20} & 1.037 \pm 0.061 & 1.587 \pm 0.040 & 0.1598 \pm 0.0059 & 0.2043 \pm 0.0033 & 4.056 \pm 0.008 & 5.023 \pm 0.006 & 3020 \pm 42\\
            \text{J0540-17} & 1.120 \pm 0.062 & 1.636 \pm 0.040 & 0.1633 \pm 0.0058 & 0.1949 \pm 0.0032 & 4.051 \pm 0.012 & 5.071 \pm 0.010 & 3180 \pm 55\\
            \text{J0546-18} & 1.051 \pm 0.059 & 1.509 \pm 0.064 & 0.2129 \pm 0.0075 & 0.2349 \pm 0.0061 & 4.085 \pm 0.016 & 5.022 \pm 0.020 & 3364 \pm 57\\
            \text{J0719+25} & 1.078 \pm 0.059 & 1.305 \pm 0.038 & 0.1584 \pm 0.0055 & 0.1917 \pm 0.0029 & 4.228 \pm 0.011 & 5.072 \pm 0.009 & 3109 \pm 59\\
            \text{J0941-31} & 1.181 \pm 0.067 & 1.745 \pm 0.046 & 0.2173 \pm 0.0078 & 0.2365 \pm 0.0036 & 4.016 \pm 0.010 & 5.025 \pm 0.008 & 3433 \pm 47\\
            \text{J0955-39} & 1.189 \pm 0.068 & 1.096 \pm 0.027 & 0.2211 \pm 0.0080 & 0.2327 \pm 0.0030 & 4.439 \pm 0.010 & 5.049 \pm 0.009 & 3300 \pm 52 \\
            \text{J1013+01} & 0.982 \pm 0.056 & 1.007 \pm 0.020 & 0.1706 \pm 0.0062 & 0.2064 \pm 0.0030 & 4.429 \pm 0.007 & 5.047 \pm 0.006 & 3028 \pm 38\\
            \text{J1305-31} & 1.063 \pm 0.059 & 1.493 \pm 0.034 & 0.2820 \pm 0.0095 & 0.2982 \pm 0.0042 & 4.133 \pm 0.010 & 4.940 \pm 0.008 & 3156 \pm 46\\
            \text{J1522+42} & 1.000 \pm 0.055 & 1.364 \pm 0.030 & 0.1656 \pm 0.0063 & 0.1915 \pm 0.0043 & 4.168 \pm 0.014 & 5.093 \pm 0.013 & 3065 \pm 49\\
            \text{J1559-05} & 1.127 \pm 0.065 & 1.709 \pm 0.037 & 0.1568 \pm 0.0058 & 0.2011 \pm 0.0058 & 4.024 \pm 0.012 & 5.025 \pm 0.019 & 3139 \pm 71 \\
            \text{J1741+31} & 1.190 \pm 0.066 & 1.187 \pm 0.023 & 0.461 \pm 0.015 & 0.377 \pm 0.018 & 4.365 \pm 0.007 & 4.948 \pm 0.042 & -- \\
            \text{J1928-38} & 0.994 \pm 0.055 & 1.384 \pm 0.028 & 0.2703 \pm 0.0091 & 0.2692 \pm 0.0057 & 4.153 \pm 0.012 & 5.010 \pm 0.009 & 3153 \pm 62\\
            \text{J1934-42} & 1.132 \pm 0.070 & 1.028 \pm 0.028 & 0.1960 \pm 0.0076 & 0.2241 \pm 0.0067 & 4.476 \pm 0.012 & 5.036 \pm 0.020 & 2982 \pm 60\\
            \text{J2040-41} & 0.997 \pm 0.055 & 1.352 \pm 0.047 & 0.1524 \pm 0.0053 & 0.1802 \pm 0.0032 & 4.170 \pm 0.013 & 5.109 \pm 0.012 & 2961 \pm 67\\
            \text{J2046-40} & 1.058 \pm 0.059 & 1.244 \pm 0.025 & 0.1917 \pm 0.0067 & 0.2212 \pm 0.0046 & 4.273 \pm 0.011 & 5.032 \pm 0.008 & 3145 \pm 41\\
            \text{J2046+06} & 1.126 \pm 0.062 & 1.608 \pm 0.032 & 0.1769 \pm 0.0062 & 0.2055 \pm 0.0025 & 4.071 \pm 0.008 & 5.060 \pm 0.006 & 3124 \pm 32\\
            \text{J2134+19} & 0.889 \pm 0.049 & 1.831 \pm 0.043 & 0.359 \pm 0.019 & 0.3706 \pm 0.0088 & 3.860 \pm 0.016 & 4.854 \pm 0.019 & 3532 \pm 43 \\
            \text{J2315+23} & 1.069 \pm 0.059 & 1.534 \pm 0.041 & 0.2309 \pm 0.0099 & 0.2521 \pm 0.0034 & 4.108 \pm 0.009 & 4.999 \pm 0.009 & 3235 \pm 51\\
            \text{J2343+29} & 1.192 \pm 0.071 & 0.914 \pm 0.017 & 0.1202 \pm 0.0046 & 0.1464 \pm 0.0031 & 4.596 \pm 0.011 & 5.191 \pm 0.008 & 2699 \pm 59\\
            \text{J2359+44} & 1.253 \pm 0.070 & 1.711 \pm 0.033 & 0.293 \pm 0.010 & 0.2978 \pm 0.0036 & 4.066 \pm 0.006 & 4.958 \pm 0.005 & 3484 \pm 45 \\
            \noalign{\smallskip}
            \hline
         \end{array}
    $
\end{table*}

\subsection{Comparison to previous studies}

As well as comparing to our \textit{TESS} analyses, we can compare our results to previous studies.
Comparing to the results we presented in EBLM VIII \citep{swayne21} we can observe a small difference in final results and uncertainty.
Given the similarity in radius ratio we believe the differences in final radius and effective temperature to be due to the different primary stellar parameters chosen.
This emphasises the importance in choosing accurate primary stellar parameters and in accounting for differences in method when performing comparison studies.
Even small changes in these quantities can result in derived results differing by a few percents, a similar effect as seen from stellar activity.
For the particular target of J0113+31 we can compare to the results of recent studies.
Our analysis of the \textit{TESS} light curve shown in \citet{swayne2020tess} show results very similar in values of $R_2/a$ with a difference of $\sim$ 50 K in effective temperature, with the two derived results showing overlapping uncertainty ranges.
Another analysis of the target by \citet{maxted21} derives almost identical M-dwarf stellar radius to our analysis but a hotter effective temperature with a difference of $\sim$ 120 K.
Our proposed cause for this difference in effective temperature that is not seen in stellar radius, is the different primary effective temperatures used by the different analyses.
With the primary effective temperature used by \citet{maxted21} being $\sim$ 100 K hotter than ours, this would result in a greater surface brightness derived for the primary star and thus a greater surface brightness for the secondary star being derived from the surface brightness ratio.
This would again emphasise the importance of accurate primary stellar parameters in photometric analyses.
Improvements in precision of these parameters will likewise see improvements in the precision of the secondary.

\begin{figure*}
    \includegraphics[width=0.95\linewidth]{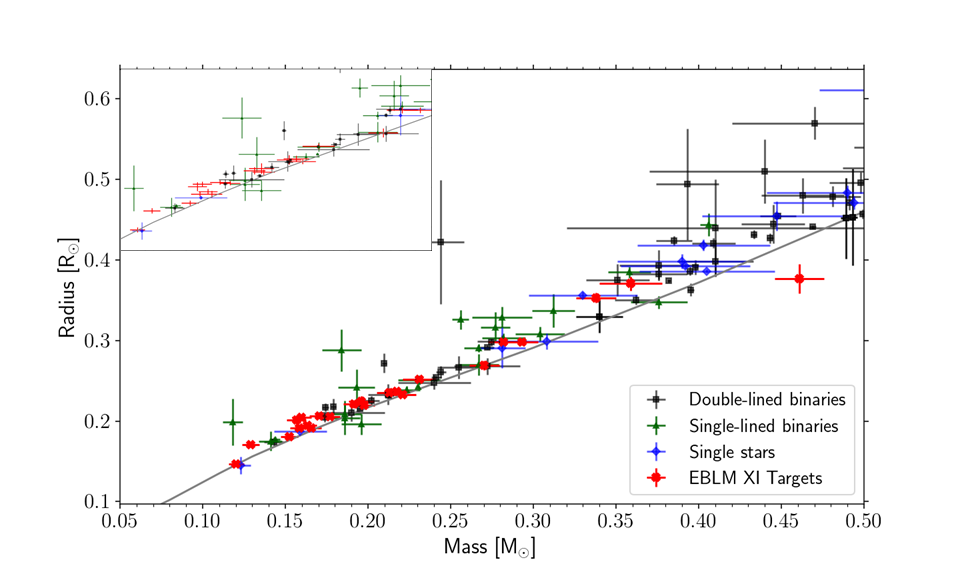} 
    \caption{A cutout of the stellar mass versus stellar radius diagram using results from \protect\citet{nefs2013,gillen,parsons2018scatter,Jennings,Duck, martinsub} with our results highlighted in red.
    The type of system is displayed by different colours. 
    The theoretical relation from \protect\citet{Baraffe15} for an age of 1 Gyr is plotted in gray.}
    \label{fig:MRfin}
\end{figure*}

\begin{figure*}
    \includegraphics[width=0.95\linewidth]{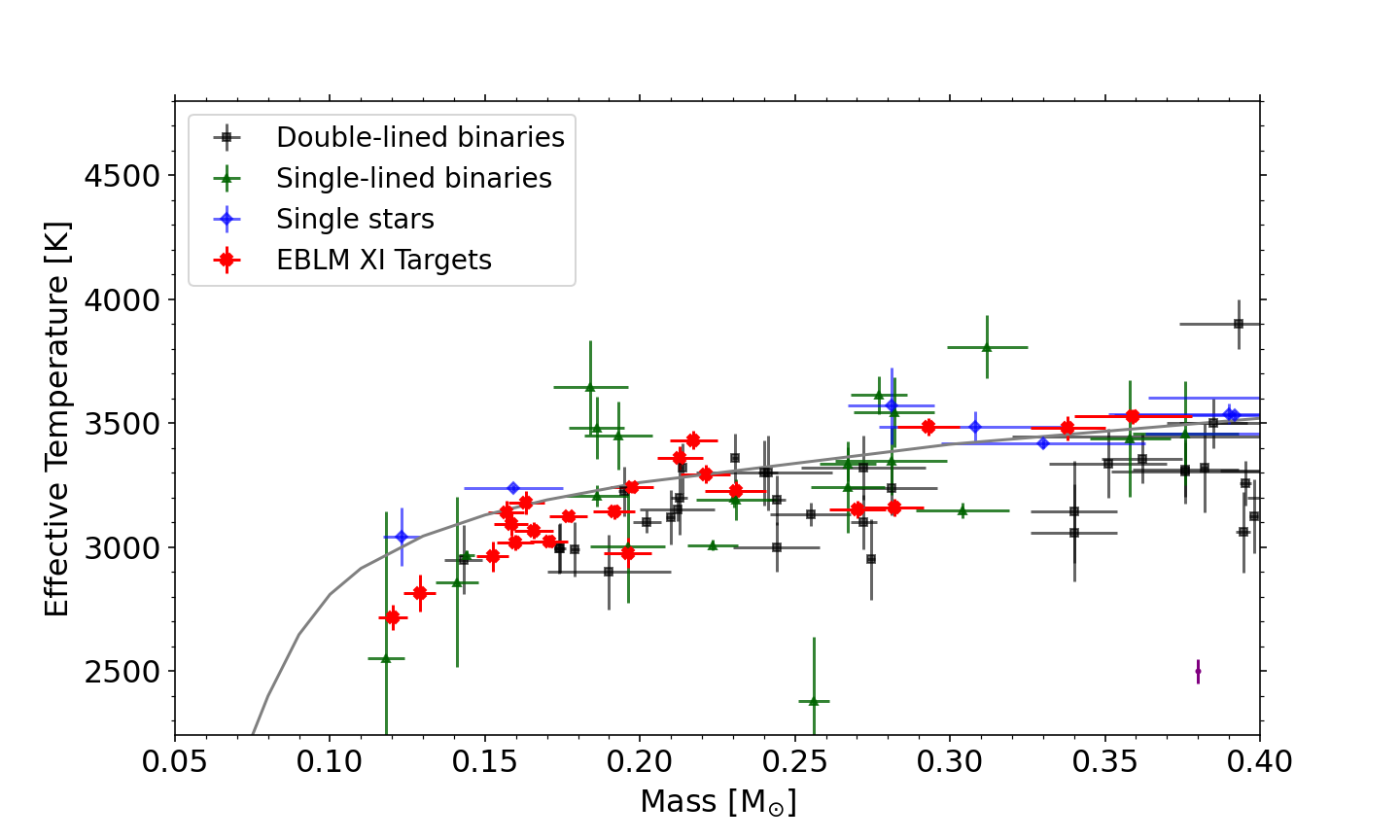} 
    \caption{A cutout of the stellar mass versus effective temperature diagram using results from \protect\citet{nefs2013,gillen,parsons2018scatter,Jennings,Duck, martinsub}, with our results highlighted in red. 
    The type of system is displayed by different colours. 
    The theoretical relation from \protect\citet{Baraffe15} for an age of 1 Gyr is plotted in gray.
    The systematic error of 50\,K that is added to our final results has been displayed in the bottom right of the Figure in purple.}
    \label{fig:MTfin}
\end{figure*}

Our final derived mass, radius and effective temperature values are shown in Figures \ref{fig:MRfin} and \ref{fig:MTfin}.
They greatly increase the number of M-dwarfs in the low-mass end of the HR diagram with both precise radii and effective temperature measurements and with known metallicity.
Our sample spans targets both in-line with the theoretical M-R and M-${\rm T}_{\rm eff}$ relations and those that seem inflated and cooler than we would expect.
This allows a thorough examination for potential causes of radius inflation .
We are also pleased to note that the precision of our derived values is in-line with or improves upon the precision of previous observations in our chosen mass range.

\section{Discussion}
\label{sec:dis}

This study assumes a uniform age for all targets. 
We make the same assumption as \citet{von2019eblm} that the evolution of stellar radii is negligible between 1 and 10 Gyr, however in further studies accounting for age would completely eliminate this as a potential factor in inflation.

\subsection{Examining potential trends with metallicity}

Metallicity was a major interest going into this project.
We sought to test the hypothesis that it is a potential cause of radius inflation, using our precise radii and metallicity calculated for us by the \textit{CHEOPS} TS3 team.
By comparing our derived radii to radii generated by theoretical structural models at the target's mass we could derive their radius inflation and search for a trend with metallicity.

We used the MIST stellar structure models \citep{Dotter2016} which can generate isochrones for metallicities up to 0.5 dex.
We download isochrones for metallicities of -0.75 to 0.5 dex (in steps of 0.25 dex) to cover the metallicity range of our targets.
From these we could draw theoretical mass-radius and mass-effective temperature relations for six metallicities and interpolate between them for the specific metallicity of the target.
Using this we obtain the theoretical radius for a given mass at the target's metallicity and derive a value for the percentage radius inflation.

\begin{figure}
    \includegraphics[width=0.95\linewidth]{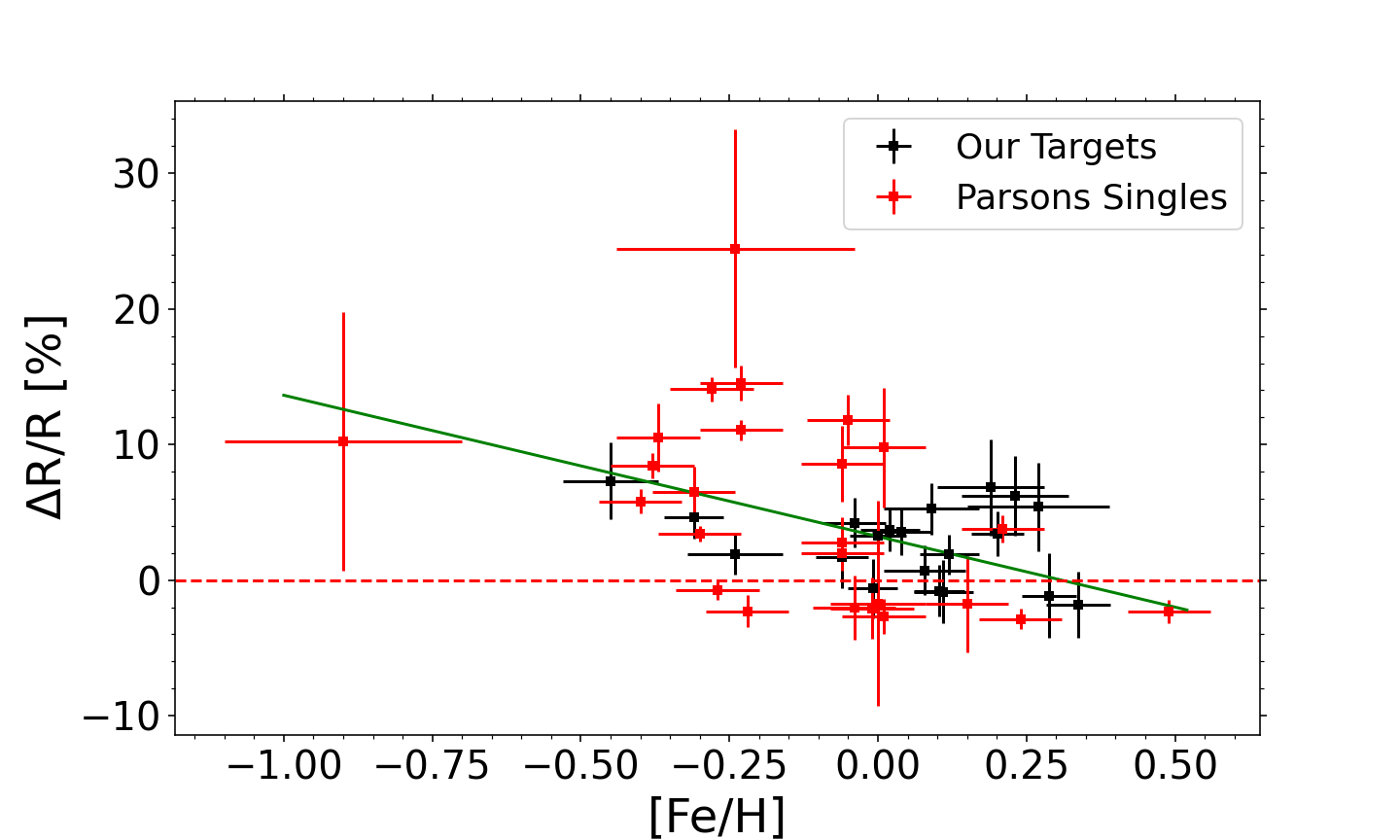} 
    \caption{The percentage radius inflation (i.e. the percentage change of our observationally derived radii from the theoretical stellar radii), versus the target's metallicity for all our targets and the single object systems in \protect\citet{parsons2018scatter}. A weighted linear fit of the data is plotted over the data in green.}
    \label{fig:inflZall}
\end{figure}

We display the metallicity versus inflation relation for single M-dwarf systems from \citet{parsons2018scatter} alongside our own targets in Figure \ref{fig:inflZall}.
Theoretical radii for each single M-dwarf was determined by interpolating in mass and metallicity using the same methods as for our own targets.
To explore any potential trend in the collected data, we performed a weighted linear fit.
A straight line polynomial was fit using the uncertainty in inflation and the scatter of the points around the straight line fit as weights.
We then adjusted the value for the point scatter until our fit produced a reduced chi-squared value of 1.
This resulted in the linear fit shown in Figure \ref{fig:inflZall}.
This fit line has a gradient of $-0.089 \pm 0.029$ i.e., $>$ 3 standard deviations difference from a zero slope, indicating a potentially significant trend between metallicity and inflation.
However we note that the majority of results are clustered around solar metallicity and that taking each sample in isolation results in different fit line gradients.
Supporting this point are the results of EBLM~V \citep{von2019eblm}, who observe a fit of the opposite trend to our own in their Figure~6, though similarly they have most results clustered around solar metallicity.
Taken in isolation the results of different studies would find entirely different relations between radius inflation and metallicity, with a negative correlation in \citet{Feiden2013b}, a positive correlation in \citet{von2019eblm} and no correlation initially in \citet{parsons2018scatter}.
With our own results in Figure \ref{fig:inflZall}, this leaves us unable to rule-out that differences between studies' methodologies could be behind such different results.
To fully explore whether a linear trend truly exists there needs to be further observations of M-dwarfs in the low and high metallicity regimes, where there are currently very few targets.
This must be done with consistency in analysis methodology and measurement of metallicity to eliminate all possible systematic differences, preferably with a re-examination of existing studies.

\begin{figure}
    \includegraphics[width=0.95\linewidth]{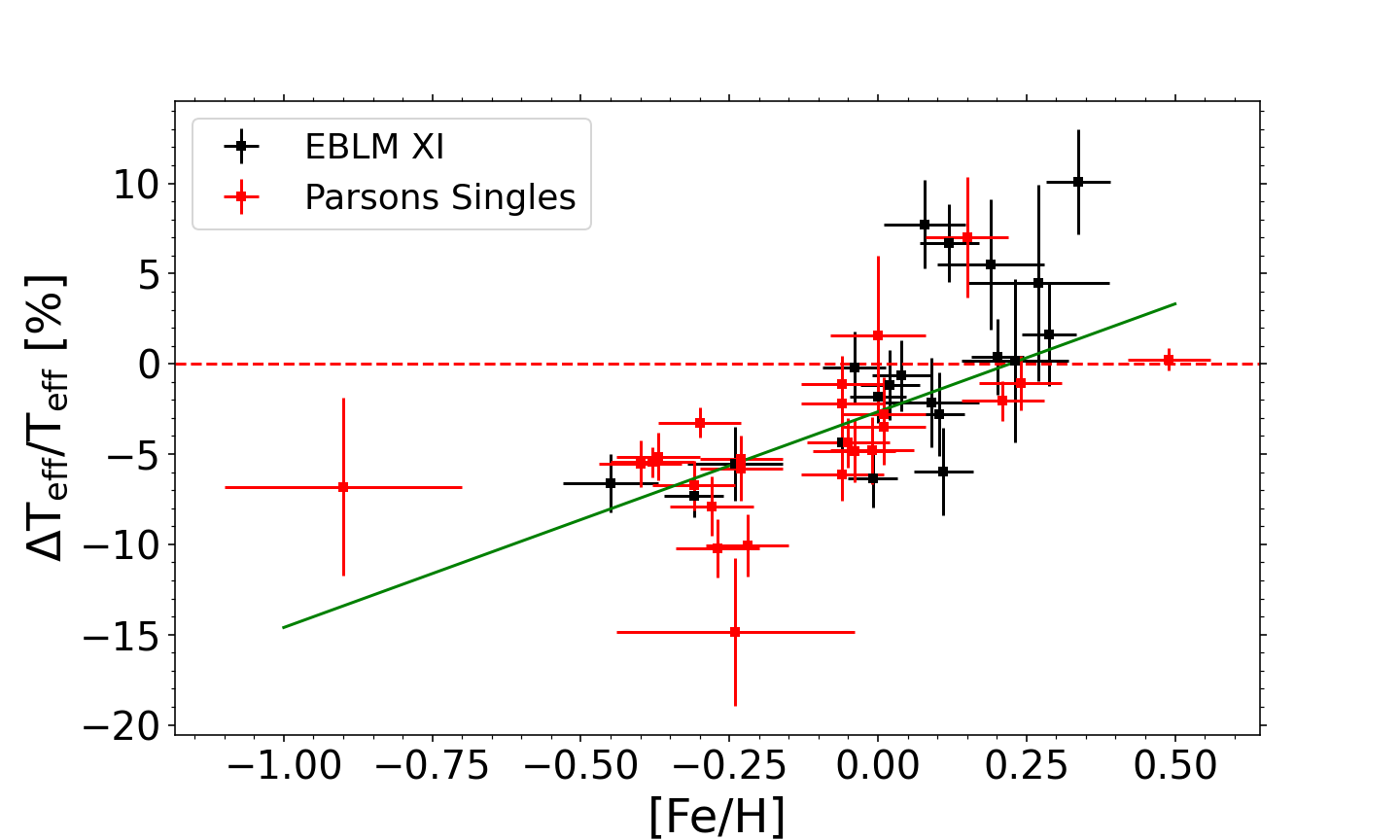} 
    \caption{The percentage effective temperature anomaly, versus the target's metallicity for all our targets and the single object systems in \protect\citet{parsons2018scatter}. A linear fit of the data is plotted over the data in green.}
    \label{fig:TinflZall}
\end{figure}

As increases in radius are theorised to come with a decrease in effective temperature (resulting in a stable luminosity), we also sought to quantify the difference between observed and theoretical effective temperature, which we shall refer to now on as the effective temperature anomaly ($\Delta {\rm T}_{\rm eff}$).
This was done with the same method as for radius: using MIST stellar structure models to generate mass-effective temperature relations.
We then interpolate through a targets mass and metallicity, before calculating the difference with our observed values.
We display the effective temperature anomaly for each target along with the anomalies calculated for the single target M-dwarfs from \citet{parsons2018scatter} in Figure \ref{fig:TinflZall}.
We perform a linear fit of the data using orthogonal distance regression.
This fit line has a gradient of $0.120 \pm 0.020$ and is displayed on Figure \ref{fig:TinflZall}.
For effective temperature we see a clearer trend (6 standard deviations difference from a zero slope) between metallicity and effective temperature anomaly then we do for radius inflation.
This trend is also present in the sample of single M-dwarf targets from \citet{parsons2018scatter} with the exception of a couple of outliers.
These results would suggest a strong correlation between effective temperature anomaly and metallicity.
With the less clear trend between radius inflation and metallicity this also would call into question the suggestion that luminosities are being measured accurately.

We use MIST due to its parameter range being compatible to our own.
Other stellar models either did not cover super-solar metallicities or did not fully cover the mass range of our targets.
However, we note that the accuracy of MIST models have been called into question at low masses.
\citet{Mann} find that MIST generated K-band magnitudes have a high sensitivity to metallicity that was not found in their observational mass-magnitude relation.
To test potential systematics caused by our choice of isochrone, we recalculated our inflation results using the DSEP stellar isochrones \citep{Dotter}.
DSEP is used as it has a compatible metallicity range with our targets as well as being covered in \citet{Mann}.
Two targets were discounted due to the publicly available isochrones we obtained not covering a low-enough mass range.
With DESP we derived theoretical radii and effective temperature through the same interpolation methods as we performed with MIST and compared these to our observed values.
In general, theoretical radii were increased at sub-solar metallicity and decreased at super-solar in comparison to MIST values, as expected given the increased metallicity dependence of MIST compared to DSEP.
This resulted in lower radius inflation values for sub-solar metallicity targets, higher inflation in super-solar metallicities and little difference for solar metallicities.
Overall for radius inflation using DSEP would result in closer agreement with our observed radii at sub-solar metallicity but less agreement at super-solar metallicity.
For our effective temperature results we saw lower theoretical effective temperatures at sub-solar metallicity and higher temperatures at solar to super-solar metallicity.
This would result in smaller sub-solar effective temperatures anomalies than when using MIST isochrones.
However, at solar metallicities all our temperature anomalies increase.
Indeed at metallicity solar and above, nearly all observed effective temperatures were cooler than the DSEP theoretical temperatures.
Thus, using our sample we do not observe a greater agreement with our observations using DSEP isochrones compared to MIST.
Our radius inflation results would suggest favouring MIST at higher metallicities, while favouring DSEP at lower metallicities.
Results in effective temperature anomaly see our observed correlation with metallicity eliminated when using DSEP, but also sees almost all our targets' effective temperatures be overpredicted in comparison with observations.
A thorough examination of these competing structure models in the context of radius inflation is not in the scope of this paper, but would be a valuable path to take for future studies.

\subsection{Trends with orbital period}

Another (much debated) potential source of radius inflation is tidal effects caused by the presence of the M-dwarf in a binary (or multiple) star system.
The closer the orbiting M-dwarf is to its companion star, the stronger tidal forces acting upon it could cause the star to spin-up.
The resultant increased magnetic activity could then inhibit its convection.
This could then cause the M-dwarf to expand, appearing to be at a greater size than what our models suggest.
This theory has seen papers support it \citep{ribas2006} and others display its shortcomings in explaining all observed inflation \citep{spada2013}.
An effect caused by being in a binary system would be a significant issue, with eclipsing binaries being one of the best means of calibrating fundamental parameters of M-dwarfs.
This trend would therefore result in binaries not being applicable to calibrating single target M-dwarfs.
As such we were keen to observe what our sample of 23 stars appeared to show, with the precision of \textit{CHEOPS} ensuring that we would accurately characterise any inflation trend with orbital period.

\begin{figure}
    \includegraphics[width=0.95\linewidth]{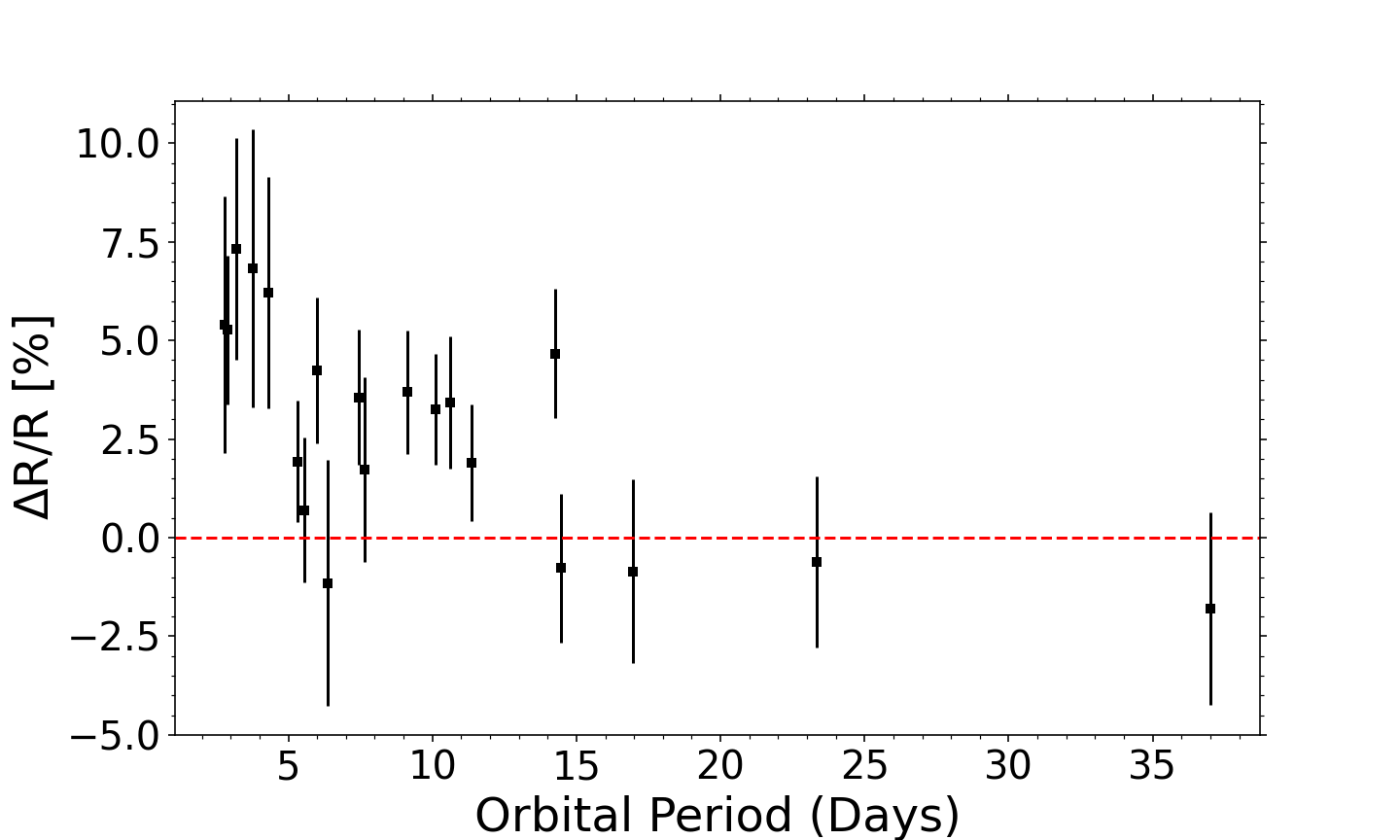} 
    \caption{The orbital period of a target versus the percentage radius inflation (i.e. the percentage change of our observationally derived stellar radii from the theoretical stellar radii).}
    \label{fig:orbPer}
\end{figure}

\begin{figure}
    \includegraphics[width=0.95\linewidth]{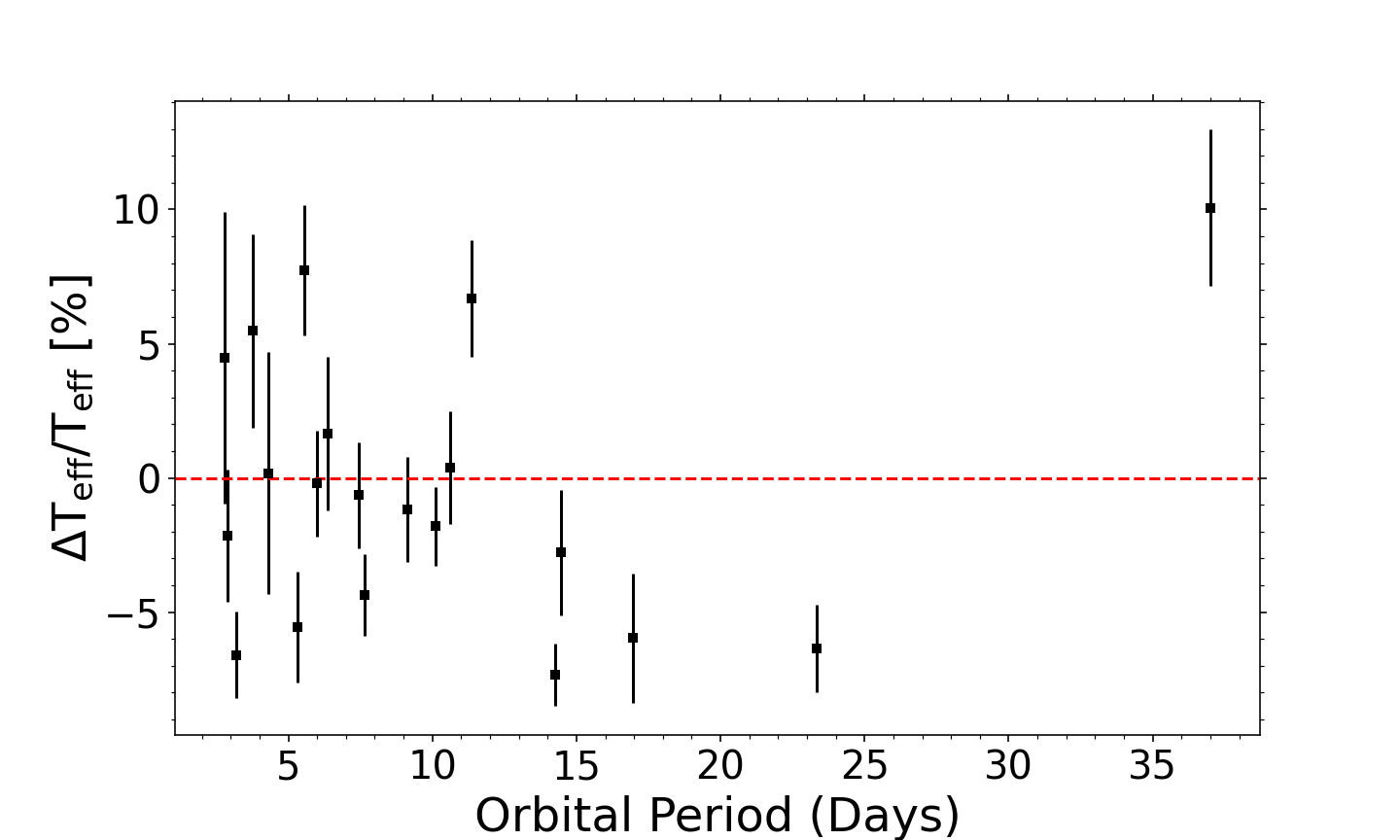} 
    \caption{The orbital period of a target versus the percentage effective temperature anomaly (i.e. the percentage change of our observationally derived stellar effective temperature from the theoretical effective temperature).}
    \label{fig:TorbPer}
\end{figure}

Our targets, shown in Figure \ref{fig:orbPer} seem to suggest a trend between orbital period, with the most inflated stars occurring at close-in orbital configurations and a lack of non-inflated values for our targets with periods lower than 5 days.
This would indicate some role for tidal forces in causing radius inflation, suggesting theoretical models need to account for these forces in the case of low mass stars in eclipsing binaries.
However, this is by no means a conclusive trend.
Our sample, although over a good range of orbital periods and separations, is relatively sparse at periods over 20 days.
Thus, we cannot conclude that our results alone definitively show a reduction in inflation with increasing orbital separation.
We also show orbital period against the effective temperature anomaly in Figure \ref{fig:TorbPer}.
The observed trend in radius inflation at low orbital separations is not reproduced in effective temperature, with there being no discernible effect on the effective temperature anomaly.
This would suggest that observed radius inflation effects could be due to being in binary systems but the ``complementary'' reduction in effective temperature is due to systematic errors in flux calculation in theoretical models.


The effect of rotation caused by tidal locking at close orbital periods on the radii of low-mass stars has been examined previously and could be a potential reason for our inflation.
This would mirror the results of \citet{kraus2011} which show inflation by the influence of a close companion.
Although this is contrasted by \citet{parsons2018scatter} who find no such link with rotation, they do also find that longer period systems appear more consistent with theoretical relations.
However, the differences between the studies leave us unable to draw perfect comparisons.
\citet{kraus2011} looks at a mass range completely different to our own, with our results focusing on very low-mass stars below the fully convective boundary (0.35 ${\rm M_\odot}$).
Nearly all of the binary targets in \citet{parsons2018scatter} do not have metallicities, meaning we cannot account for the effects observed in the previous section in our theoretical stellar radii and effective temperature.
Though it is possible the previous results between orbital period and inflation in the literature are due to systematic errors in metallicity, we cannot rule out an actual physical effect.
Therefore, deriving accurate metallicity for eclipsing binary systems is of utmost importance for existing and future observations.
\citet{von2019eblm} have derived metallicities for their targets and found no significant trend between orbital period and radius inflation, which contrasts our results.
We note that for super-solar metallicities they found it necessary to extrapolate for theoretical radii due to the lack of super-solar isochrones available in their Exeter/Lyon model grid.

\subsection{Testing the constant luminosity hypothesis}

\begin{figure}
    \includegraphics[width=0.95\linewidth]{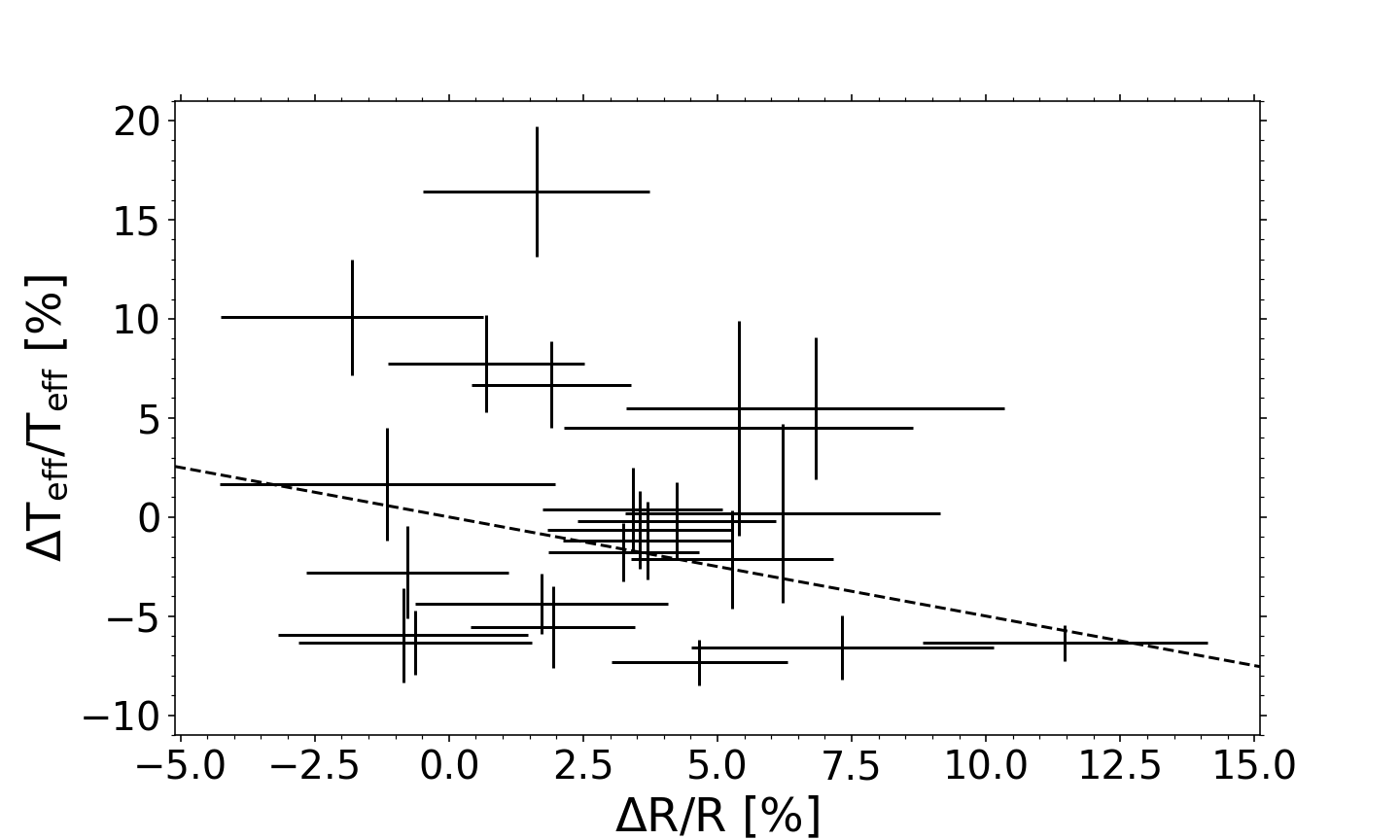} 
    \caption{The percentage radius inflation versus the percentage effective temperature anomaly. The hypothesis that these combine to leave luminosity unaffected is represented by the black-dashed line.}
    \label{fig:rinfTinf}
\end{figure}

It has been theorised that due to a correlation between radius inflation and effective temperature anomalies, luminosities predicted by models for low-mass stars are accurate \citep{Delfosse00,torres2002,ribas2006,Torres2006,Torres2007}.
This coupling of inflated radius with cooler effective temperature and vice versa has been termed the `constant luminosity hypothesis' \citep{Jennings}.
More recent measurements and derived relations suggest that radius-temperature balance is only accurate to a few \% \citep{Mann}.
As our results observe a potential decoupling between radius inflation and effective temperature anomaly, we sought to test the hypothesis.
We plot the percentage radius inflation versus the percentage effective temperature anomaly in Figure \ref{fig:rinfTinf}.
The constant luminosity hypothesis would result in a linear trend of gradient -0.5, this is shown by the black-dashed line.
Our results do not hold to this trend, with any attempted linear fitting of our results finding nothing statistically significant.

\subsection{Irradiation}
Irradiation of the M-dwarf by the primary star may play a role in radius inflation for some of the EBLM systems with the shortest orbital periods. A useful quantity to consider in this context is $F_{\rm irr} = (R_2/2a)^2 L_1/L_2$, which is the flux from the primary star intercepted by the M-dwarf relative to its intrinsic luminosity assuming a circular orbit. This quantity is $\la 2$\,per~cent for most of the stars in this sample, but is 6\,--\,10\,per~cent for three of the EBLM systems with P$<5$\,days. 
 
\section{Conclusions}
\label{sec:conc}

In EBLM XI we set out to better populate the low-mass end of the stellar H-R diagram and provide a resource to explore the effect of radius inflation for low mass stars.
In this respect its basic goal has been achieved, generating a sample of precise mass, radius and effective temperature measurements.
This well-characterised sample will act as a useful resource for further research on radius inflation, EBLMs and low mass stars in-general.
Our programme has also demonstrated the benefits of our methods of observation.
High quality photometric light curves, combined with precise radial velocity data, allows the accurate characterisation of M-dwarf stars and an exploration of their properties.
With the benefits of observing EBLMs, including using the reliable metallicity of the larger primary, we can derive precise effective temperatures and explore metallicity-dependent trends.
Going forward, our methodologies can be applied to further photometric observations of EBLMs, increasing the population of well characterised low mass stars.

In this paper we have reported potential significant trends with radius inflation and effective temperature anomaly.
When stellar metallicity is considered in calculating theoretical stellar radii, any trend between metallicity and radius inflation lessens while still being apparent (Fig.~\ref{fig:inflZall}).
This contrasts with the clearer trend between metallicity and the effective temperature anomaly (Fig.~\ref{fig:TinflZall}), though further research must be done on the role of stellar isochrones before we conclude this effect is definitive.
M-dwarfs in EBLM systems with orbital periods $<5$ days are clearly inflated compared to M-dwarfs in longer-period systems (Fig.~\ref{fig:orbPer}). 
For the stellar models we have used, the radius inflation is about 6\,per~cent, compared to about 3\,per-cent or less for longer-period EBLM systems. 
This suggests that M-dwarfs in EBLM binaries with orbital periods $<5$ days may not be suitable for testing single-star models or calibrating empirical relations to characterise planet-host stars.
There is no corresponding decrease in the effective temperature for orbital periods $<5$ days, as might be expected if radius inflation does not impact the mass-luminosity relation for M-dwarfs (Fig.~\ref{fig:TorbPer}). 
This suggests that radius inflation and effective temperature anomalies are separate phenomena.
Observation of systems with orbital periods $\ga 15$ days would be helpful to explore whether the fall-off in inflation towards higher separations seen in Fig.~\ref{fig:orbPer} is a real effect.
Low and high metallicity targets must be observed to fill-out the wings of the metallicity-inflation relation.
Furthermore, a re-examination of previous results with differing conclusions (e.g. \citealt{von2019eblm}) with our methodology would be worthwhile, ruling-out differences in methods or models used for the conflicting results.

We attempted to generate empirical relations between \textit{Gaia} magnitude $M_G$ and our results for mass, radius and effective temperature.
A tight fit to magnitude could not be achieved with a number of seemingly anomalous values.
This could potentially be due to jitter in the orbital parallaxes used in our calculations or due to a metallicity dependence at this wavelength.
For this reason, secondary eclipse measurements in the \textit{J}, \textit{H} or \textit{K} bands would be desirable as relationships between magnitude and absolute parameters have been found to have less scatter with metallicity \citep{Delfosse00,MannI}.
Our results could be used in the generation of empirical relations for mass, radius and effective temperature with these eclipse depth measurements or once improved parallaxes from \textit{Gaia} DR4 are available.
These empirical relations would then be a valuable resource for observers of low-mass stars and the exoplanets orbiting them.
In this way our work can provide not only further direction to the radius inflation problem but help guide future scientists in observing and working with low-mass stars.
With upcoming projects such as the ESA's \textit{PLATO} satellite \citep{2018PLATO}, the techniques in this paper can be used as newer and more precise instruments are focused upon EBLMs.
Low mass stars will continue to be of great interest in the coming decade and in this work we contribute towards making them a more reliable target and highlighting paths of interest for future research.